\documentclass[aps,prx,twocolumn,amsmat,amssymb,amsfonts,superscriptaddress]{revtex4-2}

\usepackage{amsmath, amsfonts}
\usepackage{bbold}

\usepackage{hyperref}
\hypersetup{colorlinks=true,
		    linkcolor=blue,
		    citecolor=blue,
		    allcolors=blue}
\usepackage{color}
\usepackage{graphics, graphicx}

\usepackage{dcolumn}
\newcolumntype{d}[1]{D{.}{.}{#1}}
\usepackage{multirow}
\usepackage{ifthen, times}
\usepackage{tikz}
\usetikzlibrary{arrows}
\usetikzlibrary{decorations.pathreplacing,decorations.markings}

\usepackage[percent]{overpic}
\usepackage{wasysym}

\DeclareMathOperator{\Tr}{Tr}
\makeatletter
\g@addto@macro\bfseries{\boldmath}
\makeatother

\usepackage{bm, bbm, braket, mathtools, comment}
\usepackage{tcolorbox}
\tcbuselibrary{theorems}
\renewcommand{\H}{\mathcal{H}}

\newcommand{\Minus}{\mathord{\tikz\draw[line width=0.3ex, x=1ex, y=1ex] (0.0,1.0) -- (1.,1.0);}}
\newcommand{\xbond}{{\color{blue}\Minus}}
\newcommand{\ybond}{{\color{red}\Minus}}

\newcommand{\shalf}{$S$=$\tfrac{1}{2}$}

\newcommand{\sone}{$S$=1}
\newcommand{\stwo}{$S$=2}

\begin{document}

\title{Islands of Chiral Solitons in Integer Spin Kitaev Chains}
\author{Erik S. S{\o}rensen}
\email{sorensen@mcmaster.ca}
\affiliation{Department of Physics \& Astronomy, McMaster University, Hamilton ON L8S 4M1, Canada.}
\author{Jonathon Riddell}
\email{Jonathon.Riddell@nottingham.ac.uk}
\affiliation{Department of Physics \& Astronomy, McMaster University, Hamilton ON L8S 4M1, Canada.}
\author{Hae-Young Kee}
\email{hykee@physics.utoronto.ca}
\affiliation{Department of Physics, University of Toronto, Ontario M5S 1A7, Canada}
\affiliation{Canadian Institute for Advanced Research, CIFAR Program in Quantum Materials, Toronto, ON M5G 1M1, Canada}
\date{\today}

\begin{abstract}
An intriguing chiral soliton phase has recently been identified in the \shalf\
Kitaev spin chain. Here we show that for $S$=1,2,3,4,5 an analogous phase can
be identified, but contrary to the \shalf\ case the chiral soliton phases
appear as islands within the sea of the polarized phase. In fact, a small
field applied in a general direction will adiabatically connect the integer
spin Kitaev chain to the polarized phase. Only at sizable intermediate fields
along symmetry directions does the soliton phase appear centered around the
special point $h^\star_x$=$h^\star_y$=$S$ where two {\it exact} product
ground-states can be identified. The large $S$ limit can be understood from a
semi-classical analysis, and variational calculations provide a detailed
picture of the \sone\ soliton phase. Under open boundary conditions, the
chain has a single soliton in the ground-state which can be excited, leading
to a proliferation of in-gap states. In contrast, even length periodic chains
exhibit a gap above a twice degenerate ground-state. The presence of solitons
leaves a distinct imprint on the low temperature specific heat.
\end{abstract} 
\maketitle

\section{Introduction}\label{sec:intro}
Shortly after a microscopic mechanism to realize the exactly solvable \shalf\ Kitaev model defined on the two-dimensional honeycomb lattice\cite{kitaev2006} was proposed\cite{jk2009prl}, intense research in generalizations of Kitaev's original model started,
including other interactions, higher-spin models, and/or external magnetic field.
From a materials perspective, Kitaev materials, broadly defined as materials with dominant bond-dependent interactions, possess surprisingly rich and intricate phase diagrams~\cite{balents2014review,rau2016review,winter2017review,hermanns2018review,Janssen2019review,Takagi2019review,trebst2022review}.
Notably, in the presence of an applied field, 
Kitaev models
lead to a phase diagram not only depending on field strength but also on field direction, with a resulting proliferation of competing phases. Of particular interest are field-induced spin liquid phases, where intriguing results been suggested in recent experiments on the \shalf\ material $\alpha$-RuCl$_3$ when an in-plane field~\cite{Kasahara2018,Yokoi2021,Czajka2021,Bruin2022,Czajka2022}
or out-of-plane field~\cite{Zhou2022} is applied. 
In theoretical studies of \shalf\  antiferromagnetic (AFM) Kitaev honeycomb models, signatures of possible spin liquid phases under a magnetic field have also been reported.~\cite{zhu2018prb,nasu2018prb,liang2018prb,gohlke2018prb,lu2018spinon,hickey2019visons,patel2019,zou2020neutral}
Near the ferromagnetic (FM) Kitaev regime, a field-induced intermediate phase was found when the magnetic field is at or close to the out-of-plane direction~\cite{Gordon2019,kaib2019prb,Lee2020Magnetic,Li2021}.

Another focus has been higher spin Kitaev models with $S>\frac{1}{2}$~\cite{Baskaran2008,Rousochatzakis2018}. Initially an academic problem,  a microscopic theory showed that  utilizing Hund's coupling in transition metal cations and spin-orbit coupling at anions led to a higher-spin Kitaev interaction~\cite{Stavropoulos2019}. In particular, \sone\ models~\cite{Koga2020,Dong2020,Zhu2020,Khait2021,Chen2022} where the presence of a gapless spin liquid phase for AFM Kitaev model at finite field
has been suggested~\cite{Hickey2020}.
While these field-induced magnetically disordered phases in \shalf\ and higher-S are fascinating, the precise nature of these phases and the physical mechanisms giving rise to them is still not completely understood. One challenge is associated with the size of the systems that one can access in numerical studies.
\begin{figure}
    \includegraphics[width=\columnwidth]{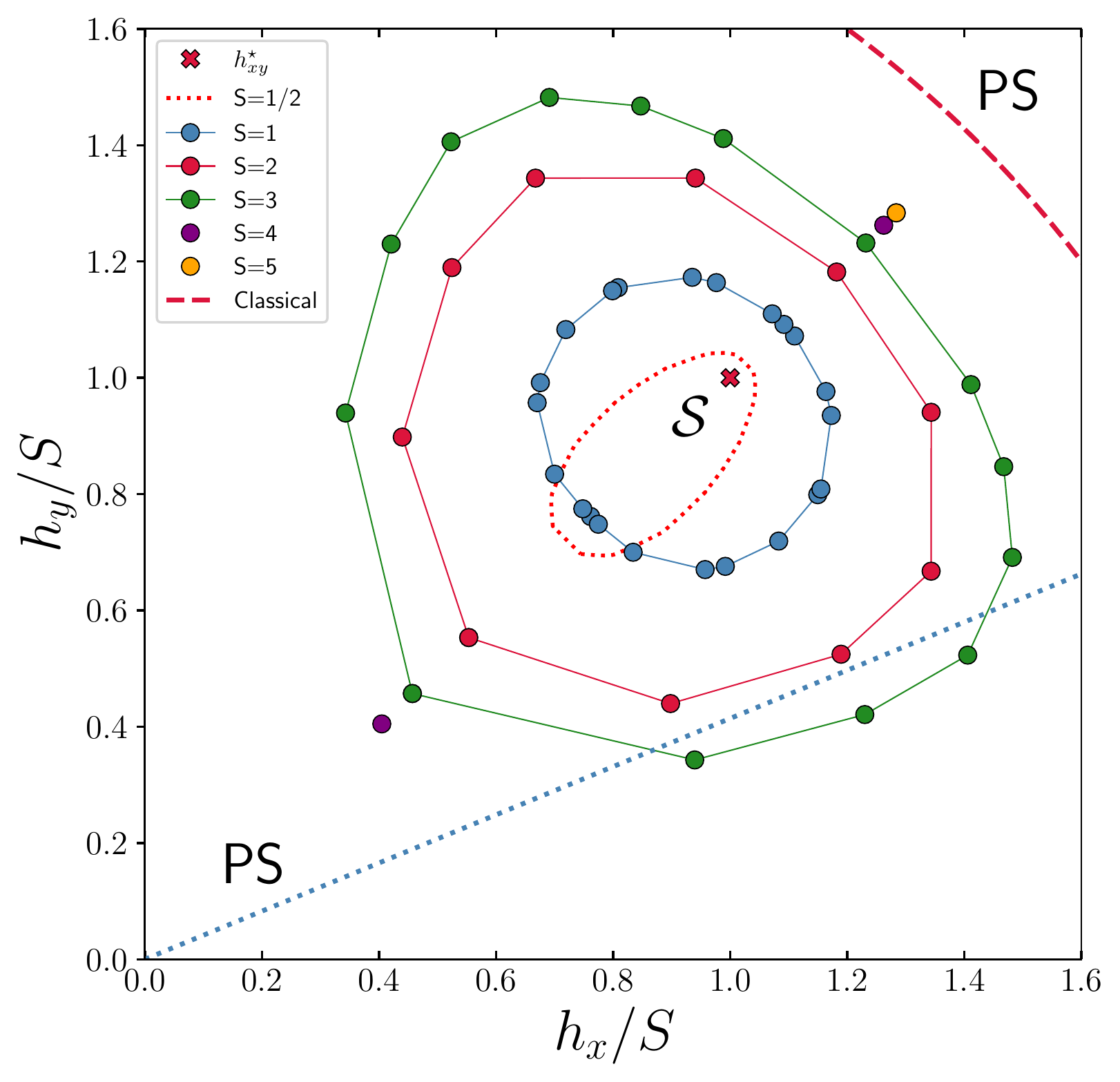}
    \caption{
    iDMRG results for the \sone(blue), 2 (red), 3 (green), 4 (purple), 5 (orange) Kitaev spin chain. Points indicate peaks in
    $\chi^e_h$ or $\chi^e_{\phi_{xy}}$. The dashed red line indicate the classical value for the transition to the polarized state, and the dotted red line are results for \shalf\ from Ref.~\cite{Sorensen2022}. The red cross indicates $h^\star_{xy}$=$S\sqrt{2}$ with $h^\star_x/S=h^\star_y/S=1$.
    }  
    \label{fig:phasediagram}
\end{figure}

To gain insight into the field-induced phases, a different route was recently taken, instead starting with low-dimensional versions of the Kitaev model such as chains and ladders under a magnetic field where highly precise results can be obtained for very large systems or in the thermodynamic limit.
While geometrically restricted, interesting
chiral phases near AFM Kitaev region in a perpendicular field have been identified~\cite{Sorensen2021}
in \shalf\ ladder models.
An extended soliton phase induced by the field in the \shalf\ Kitaev spin chain was also recently discovered ~\cite{Sorensen2022}. 

An early work by Sen et al.~\cite{Sen2010} showed that the spin-$S$ Kitaev chains have an analog
of the Z$_2$ conserved quantities present in Kitaev's honeycomb model and demonstrated that there is a 
qualitative difference between the integer and half-integer spin
due to their different commutation relations. They also showed that  the
\sone\ chain exhibits a unique ground state with local excitations
of the Z2 conserved quantities, which was later confirmed by numerical studies~\cite{Luo2021}.
It is then natural to ask if the field-induced soliton phase arise in 
Kitaev spin chains with integer spins, a question we answer in the affirmative here. 

The rest of the paper is organized as follows.
We present the model Hamiltonian and a main result of the phase diagram in field strength and direction in the next section. 
In section~\ref{sec:num} we briefly summarize the main numerical techniques we have used.
Section~\ref{sec:results} presents our iDMRG and DMRG results used for determining the phase diagram, excitation gaps, chiral ordering as well as the soliton mass and size.
Section~\ref{sec:classical} describes the uniform product states approximating the two ground-states within the soliton phase for any $S$ with periodic boundary conditions.
A variational picture based on previous results for the \shalf\ model in Ref.~\cite{Sorensen2022} is then developed in
In section~\ref{sec:variational}, and penultimately we discuss in section~\ref{sec:specheat} how signatures of the solitons can
be detected in the specific heat, in particular for open boundary conditions.
Finally, in section~\ref{sec:Conclusion} we present a discussion of our results and remaining open problems.

\section{Model, Phase Diagram and Phenomenology}
The Kitaev spin chain is described by the Hamiltonian:
\begin{equation}\label{eq:H}
\H = K\sum_{j}\left(S_{2j+1}^xS_{2j+2}^x + S_{2j+2}^yS_{2j+3}^y\right)- \sum_j \bm{h} \cdot \bm{S}_j,
\end{equation}
where we set $g$=$\hbar$=$\mu_B$=$1$ and consider the AFM model with $K$=1. Furthermore, we  parameterize the field term as
$\bm{h}$=$h(\cos\phi_{xy}\cos\theta_z,\sin\phi_{xy}\cos\theta_z,\sin\theta_z)$ and define $|\mathbf{h}|$ as the field strength. We use $N$ to denote the number of sites in the model, and we shall refer to the $KS^xS^x$ coupling as a $x$-bond ($\xbond$) and the $KS^yS^y$ coupling as a $y$-bond ($\ybond$).
The $S>\frac{1}{2}$ Kitaev chain was considered in Ref.~\cite{Sen2010} and the \sone\ model in zero field
has been the subject of several studies~\cite{Liu2015,You2020,Luo2021,You2022}, however, to our knowledge the phase diagram
in the presence of a magnetic field has not previously been investigated, likely since it has been assumed that 
the model would transition to the polarized phase without  any intervening non-trivial phases as has been shown 
to be the case for the \shalf\ chain in a transverse magnetic field~\cite{Sun2009}. However, it turns out that if more general
field directions are considered a highly non-trivial soliton phase can be identified in the \shalf\ chain~\cite{Sorensen2022}, appearing along the field directions $\phi_{xy}$=$\frac{\pi}{4}$+$n\frac{\pi}{2}$.

As we show in section~\ref{sec:phenom} and \ref{sec:phasediag}, for integer $S$, the soliton phase appears as an unusual {\it reentrant} island arising out of the sea
of the polarized state (PS). If the magnetic field already has forced the chain to enter the polarized phase,  the appearance of a non-trivial soliton phase as the magnetic field is further increased may at first sight seem counterintuitive. However, at the unique field strength $h^\star_{xy}$ we identify two {\it exact} ground-states for any $S$ with periodic boundary conditions (PBC), which allows us to develop variational arguments showing that such a soliton phase indeed must exist in the vicinity of $h^\star_{xy}$. Furthermore, the existence of such a soliton phase appears to rely on the presence of a gap for periodic boundary conditions, while open boundary conditions should give rise to numerous in-gap states.
We mainly focus on the integer spin case since the low field physics of the half integer spin chains is subtly different~\cite{Sorensen2022} but we expect many of our results, in particular the existence of the soliton phase, to be valid for any $S$.

Our main results for the phase diagram of the integer spin Kitaev chain, Eq.~(\ref{eq:H}), are summarized in Fig.~\ref{fig:phasediagram} where the soliton phase is shown in the first $h_x,h_y$ quadrant for \sone, 2, 3, 4 and 5. By symmetry, a similar phase diagram applies to the other 3 quadrants in the $h_x, h_y$ plane with $\phi_{xy}$=$\frac{\pi}{4}$+$n\frac{\pi}{2}$. As discussed in section~\ref{sec:classical}, in the classical limit we expect solitons to be present for any $h_{xy}/K<2S$ along the line $h_x=h_y$ and the fact that the size of the soliton phase is growing with $S$ is consistent with this. On the other hand, it is clear that the soliton phase shrinks as $S$ is decreased. Surprisingly, as was shown in~\cite{Sorensen2022}, it 
survives in the \shalf\ limit as indicated in Fig.~\ref{fig:phasediagram} by the dotted red line.

Solitons in spin chains have been studied from the late seventies starting with the work of Mikeska~\cite{Mikeska1978,Mikeska1980} and Fogedby~\cite{Fogedby1980a,Fogedby1980b} and several reviews and monographs are now available~\cite{Mikeska1991,Kosevich1990,Vachaspati,Solitons}. At the same time, solitons in conducting polymers have been investigated~\cite{Heeger1988}. Initially, classical ferromagnetic (FM)
models with an easy-axis Ising symmetry were considered, where  two equivalent ground-states can be identified.
It is then straightforward to see that domain walls can be formed between the ground-states which should be regarded
as topological solitons linking distinguishable ground-states~\cite{Solitons} as opposed to hydrodynamic or non-topological solitons that cannot exist at rest~\cite{Solitons}.
In the continuum approximation, the sine-Gordon model is then applicable, leading to the well known kink solutions describing the domain walls. Experiments on the 1D easy-plane ferromagnetic chain system CsNiF$_3$~\cite{Kjems1978} confirmed
the presence of solitons and subsequent studies of 1D anti-ferromagnetic materials TMMC~\cite{Boucher1985,Regnault1982}, CsCoBr$_3$~\cite{Buyers1986,Braun2005} and CsMnBr$_3$~\cite{Gaulin1985,Gaulin1987a,Gaulin1987b} also validated the existence of solitons excitations.
Domain walls between degenerate ground-states in dimerized spin chains, such as the \shalf, $J_1$-$J_2$ model, have also
been viewed as solitons~\cite{Shastry1981,Caspers82,Caspers84,Sorensen1998,Sorensen2007a,Sorensen2007b} and observed experimentally in BiCu$_2$PO$_6$ above a critical field~\cite{Casola2013} as well as in CuGeO$_3$~\cite{Horvatic1999}. 
However, in all cases one associates a {\it positive} mass, $\Delta_s>0$,  with the soliton which appear as an {\it excitation} above the ground-state and never as the unique ground-state as we find here.
One might argue against this on the grounds that 
for $N$ odd a single soliton is always present in the dimerized chains, however, the energy is still higher than the comparable even $N$ system
indicating a positive mass of the soliton.

Before turning to a detailed presentation of our results in section~\ref{sec:results}, \ref{sec:classical}, \ref{sec:variational} and \ref{sec:specheat} it is useful to give a largely phenomenological overview of the central
mechanism and physics behind the soliton phase which we do in the following.

\subsection{Phenomenological Description of the Soliton Phase}\label{sec:phenom}
At the phenomenological level, we may understand the appearance of the soliton phase along the $h_x$=$h_y$ field direction in the following way. At high fields, all the spins align with the field, and we are in the polarized state (PS). Since the spins on all the bonds are aligned in a parallel manner, there is  a large energy cost arising from the Ising Kitaev terms on each bond that has to be overcome to sustain the polarized state. As the field is lowered the Zeeman term is not enough to overcome this energy cost, instead the chain enters one of the following two product states
\begin{equation}
    |XY\rangle=|xyxy\ldots\rangle,\ \ |YX\rangle=|yxyx\ldots\rangle.
\end{equation}
Here $|x\rangle$ and $|y\rangle$ refer to eigen-states of $S^x$ and $S^y$ and $|XY\rangle$ is shorthand for the state with $|x\rangle$ on odd sites and $|y\rangle$ on even sites. These two degenerate states are selected because
the contribution to the energy from the Kitaev terms is identically zero. On the other hand, the spins are still partially aligned with the field so the Zeeman term lowers the energy. Crucially, as we discuss further in section~\ref{sec:classical}, the $|XY\rangle$ and $|YX\rangle$ are {\it exact ground-states} for the chain at a field  
$h_x^\star$=$h_y^\star$=$KS$ and consequently $h^\star_{xy}$=
$SK\sqrt{2}$ for {\it any} $S$ under periodic boundary conditions (PBC) with energy $-NKS^2$ as long as $N$ is even as dictated by the two site unit cell. This follows from the fact that at $h^\star_{xy}$ the Hamiltonian, Eq.(\ref{eq:H}), can be written in the following form:
\begin{flalign}\label{eq:Hstar}
\ \ &\H = \H_p -NKS^2\nonumber &\\
\ \ &\H_p = K\sum_{j}\big[\left(S-S_{2j+1}^x\right)\left(S-S_{2j+2}^x\right) +\nonumber &\\
&\hskip 3cm \left(S-S_{2j+2}^y\right)\left(S-S_{2j+3}^y\right)\big].
\end{flalign}
From the form of Eq.~(\ref{eq:Hstar}), it is clear that $|XY\rangle$ and $|YX\rangle$ are the only eigen-states of $\H_p$ with an eigenvalue of zero. Furthermore, $\H_p$ is positive semidefinite proving that $|XY\rangle$ and $|YX\rangle$ are ground-states. The field value
$h^\star_{xy}$ is indicated as a green dotted line in Figs.~\ref{fig:VarSpinGap},~\ref{fig:gaps},\ref{fig:chiralZ}, \ref{fig:solitonmass}. At other field strengths $h_{xy}\neq h^\star_{xy}$, within the soliton phase, the two-fold degeneracy of the ground-state remain exact even for finite $N$ but the degenerate states are now distorted from the simple $|XY\rangle$ and $|YX\rangle$ forms.

\subsubsection{Open Boundary Conditions, Soliton Mass, $\Delta_b$}\label{sec:varsolmass}
Let us now consider the case of open boundary conditions (OBC) where the first bond is a $x$-bond ($\xbond$). We want to see if
there are other simple product states with even lower energy than the $|XY\rangle$ and $|YX\rangle$ states that can be considered with OBC. To that end, we consider states of the form
\begin{equation}
    |\psi_b(i)\rangle=|YX\ldots\nearrow_i\ldots XY\rangle,
\end{equation}
transitioning from 
$|y\rangle$ on odd  and $|x\rangle$ on even sites to the opposite pattern at site $i$ where the spin is aligned with the field, thereby maximizing the Zeeman term at that site. We then need to consider what happens to the Kitaev terms neighboring the $\nearrow$ defect. There are two possibilities:
\begin{equation}
    |\psi_b(i)\rangle=|
    y\xbond\ x\ybond\ 
    \tcbset{colback=blue!10!white}
    \tcboxmath[size=fbox,auto outer arc, arc=5pt]{
    y\xbond\ 
    \nearrow_i\ybond\ x
    }
    \xbond\ y\ybond\
    x\xbond\ y\ybond\
    x\xbond\ y\rangle,\label{eq:psib00}
\end{equation}
and 
\begin{equation}
    |\psi_b(i)\rangle=|
    y\xbond\ x\ybond\ 
    y\xbond\ 
    \tcbset{colback=blue!10!white}
    \tcboxmath[size=fbox,auto outer arc, arc=5pt]{
    x\ybond\
    \nearrow_i\xbond\ y
    }
    \ybond\
    x\xbond\ y\ybond\
    x\xbond\ y\rangle,
    \label{eq:psib01}
\end{equation}
we immediately see that due to the highly bond dependent interaction and the fact that the chain starts with an $x$-bond ($\xbond$), the energy cost of the two bonds neighboring the defect continue to be zero,
since  the $\nearrow_i\ybond\ x$ occurs on an $y$-bond with $S^y$ acting on $|x\rangle$ yielding zero and the $y\xbond\nearrow_i $ on a $x$-bond with $S^x$ acting on $|y\rangle$.
The $\psi_b$ state therefore lowers the energy with respect to the $|YX\rangle$ state without incurring an energy penalty. 
We emphasize that this effect applies equally well to odd and even $N$.
A state such as $\psi_b$, transitioning between two ground-states, is a typical example of a topological soliton linking distinguishable ground-states~\cite{Rajaraman,Vachaspati,Solitons}. One may consider other forms than the states Eq.~(\ref{eq:psib00}) and Eq.~(\ref{eq:psib01}) for the transition between the two ground-states, and in Ref.~\cite{Sorensen2022} we considered conceptually simpler bond defects which are convenient for $S$=1/2. However, since all such states are non-orthogonal, this only leads to minor differences in the final results.

Having successfully found a low-energy product state with a single defect, it is natural to consider two defects. However, if the defects are on neighboring sites, $\nearrow_i\nearrow_{i+1}$, it is clear that a large energy cost is associated with the $[i,i+1]$ bond since the spins are aligned across an antiferromagnetic bond. A second defect therefore needs to be separate from the first, creating
a transition back to the $YX$ pattern. In order to gain intuition about such a transition, let us consider 'anti-defect' states
of the form
\begin{equation}
    |B\rangle=|XY\ldots\nearrow_i\ldots YX\rangle,
\end{equation}
transitioning from
$|x\rangle$ on odd  and $|y\rangle$ on even sites to the opposite pattern at site $i$ where the spin is aligned with the field.
As before, such a state lowers the energy by aligning the spin with the field at site $i$. However, something rather extraordinary happens when we consider the bond dependent Kitaev terms neighboring this anti-defect. They can take one of the two generic forms
\begin{equation}
    |\psi_B(i)\rangle=|
    x\xbond\ y\ybond\ 
    \tcbset{colback=red!10!white}
    \tcboxmath[size=fbox,auto outer arc, arc=5pt]{
    x\xbond\ 
    \nearrow_i\ybond\ y
    }
    \xbond\ x\ybond\
    y\xbond\ x\ybond\
    y\xbond\ x\rangle,\label{eq:psiB00}
\end{equation}
and
\begin{equation}
    |\psi_B(i)\rangle=|
    x\xbond\ y\ybond\ 
    x\xbond\ 
    \tcbset{colback=red!10!white}
    \tcboxmath[size=fbox,auto outer arc, arc=5pt]{
    y\ybond\
    \nearrow_i\xbond\ x
    }
    \ybond\
    y\xbond\ x\ybond\
    y\xbond\ x\rangle,\label{eq:psiB01}
\end{equation}
in this case transitioning from the $XY$ to the $YX$ pattern at bond $i$. 
However, in this case the anti-defect incurs a high energy penalty from the Kitaev terms  
since  the $y\ybond\nearrow_i$ now occurs on a $y$-bond and the $\nearrow_i\xbond\ x'$ on a $x$-bond.
Remarkably, we see that if the chain starts with a $x$-bond, there is no way to introduce an anti-defect from $|YX\rangle$ to $|XY\rangle$ without incurring a large energy penalty. On the other hand, a single {\it defect} from $|YX\rangle$ to $|XY\rangle$ clearly lowers the energy. It follows that in the ground-state with OBC a single soliton is present and the presence of several spatially separated solitons is enegertically prohibited. However,  as we discuss in section~\ref{sec:variational} excited states of a single soliton exists leading to a proliferation of low-lying excitations.

We note that, starting the chain with a $y$-bond ($\ybond$) with a defect, transitioning from the $|XY\rangle$ to the $|YX\rangle$ pattern merely interchanges the roles of $\psi_b$ and $\psi_B$. Furthermore, the $\psi_b$ and $\psi_B$ states are not eigen-states of the Hamiltonian but, considering all possible states of the form, $|\psi_b(i)\rangle$ leads to a good description of the low-energy subspace for OBC. In section~\ref{sec:variational} we discuss variational calculations within such a subspace, and for clarity we reserve the name 'soliton' for linear combinations of the states $\Psi_b$=$\sum a_i |\psi_b(i)\rangle$. For OBC, within such a variational subspace, we can then determine by how much the presence of the soliton lowers the energy
with respect to the $|YX\rangle$ state, which we define as the soliton mass, $\Delta_b$. From the above, we expect that within the soliton phase,
\begin{equation}
\Delta_b<0,
\end{equation}
otherwise the ground-state would not be a single soliton state.
On the other hand, the $|\psi_B(i)\rangle$ are high energy states that in isolation presumably are of little relevance.
However, it is still very useful to consider linear combinations $\Psi_B$=$\sum c_i |\psi_B(i)\rangle$ thereby estimating the energy cost of an anti-soliton. In an analogous manner we can then define the anti-soliton mass $\Delta_B$ and, within the soliton phase, we expect
$\Delta_B>0$, reflecting the energy cost associated with the anti-soliton.
Even though a state such as $\Psi_B$ is not expected to be close to an eigen-state, $\Delta_B$ should still be a good estimate of the energy cost of an anti-soliton and soliton anti-soliton $bB$ states could be of low-energy and therefore relevant for periodic boundary conditions which we discuss next.

\subsubsection{Periodic Boundary Conditions - Spin Gap}\label{sec:pbc}
If we now consider periodic boundary conditions (PBC) it is clear that excitations out of the $|XY\rangle$, $|YX\rangle$ states must involve both a defect and anti-defect which we refer to as $bB$ states. Another remarkable feature of the soliton phase in the Kitaev chain is that there is no symmetry relation between the defect and anti-defect. In other systems where related physics can be observed such as the dimerized phase of the \shalf, $J_1$-$J_2$ model, where
\shalf 
domain walls between degenerate ground-states have been viewed as solitons~\cite{Shastry1981,Caspers82,Caspers84,Sorensen1998,Sorensen2007a,Sorensen2007b}, the soliton and anti-soliton are effectively indistinguishable and both {\it raise} the energy and both carry a spin of \shalf. Here, the opposite is true, the defect and anti-defect are clearly distinguishable with the defect {\it lowering} the energy while the anti-defect {\it raises} the energy ($\Delta_b<0$, $\Delta_B>0$). The defect and anti-defect are also not eigen-states of the spin operators and a definite spin cannot be associated, and we cannot ascribe the presence of the soliton to an unpaired spin. Furthermore, it turns out that the anti-defect raises the energy {\it more} than the defect lowers it. If we now imagine a defect and anti-defect well enough separated in a periodic system that their interaction can be neglected, this asymmetry in the energy cost then leads to a spin-gap above the two degenerate ground-states.
Even though the anti-defect is rather costly, the combination of the defect and anti-defect has a much smaller energy cost, creating a modest spin-gap. Not surprisingly, the maximum of the spin-gap appears to coincide with $h^\star_{xy}$ where the $|XY\rangle$ and $|YX\rangle$ product states are exact ground-states. In fact, it is clear that we must have:
\begin{equation}
\Delta_b+\Delta_B>0, \label{eq:soldef}
\end{equation}
within the soliton phase, and we can take $\Delta_b+\Delta_B$ to be a first approximation to the spin gap for PBC. Consider the opposite to be true, in that case for OBC a state with $bBb$ would have lower energy than $b$, and $bBbBb$ even lower energy, leading to a contradiction. 
Eq~(\ref{eq:soldef}) may therefore be seen as providing an estimate of the extent of the soliton phase.

\subsubsection{Critical Fields, $h_{xy}^{c1}$, $h_{xy}^{c2}$}\label{sec:hc}
\begin{figure}
  \includegraphics[width=\columnwidth]{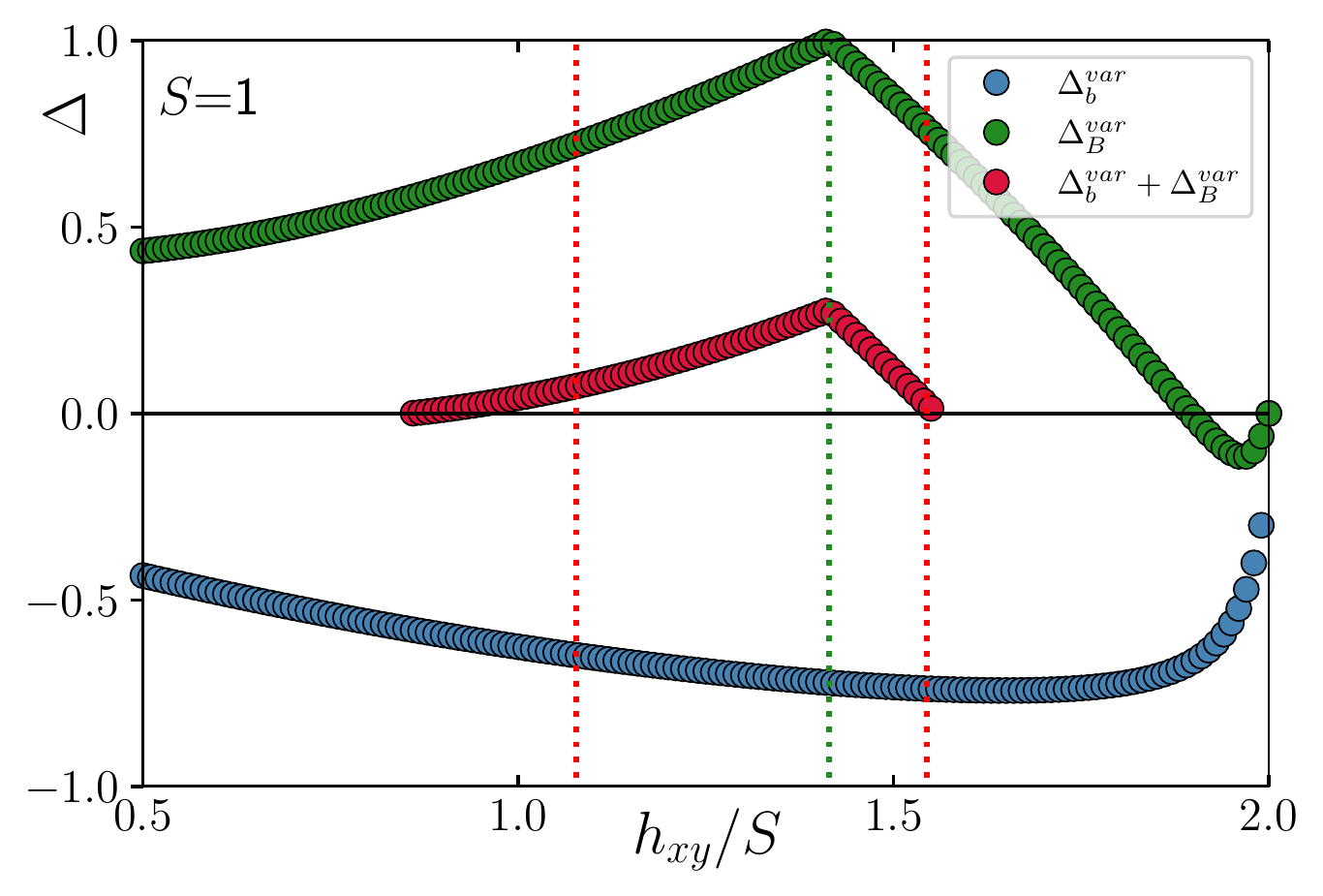}
  \caption{
  Variational estimates with OBC, $N$=$100$ and \sone\ of the soliton mass, $\Delta_b$, and anti-soliton mass $\Delta_B$ as a function of field $h_{xy}$ shown with the resulting estimate of the spin gap, $\Delta_b+\Delta_B$. Only for a finite range of fields is the spin gap positive and the  soliton phase stable. The dotted red lines are the critical fields $h_{xy}^{c1}$ and $h_{xy}^{c1}$ obtained from iDMRG, the green dotted line is $h^\star_{xy}$.
  }
  \label{fig:VarSpinGap}
\end{figure} 
As we shall discuss further in section~\ref{sec:variational},
for fields $h_{xy}\neq h^\star_{xy}$ the states $|XY\rangle$ and $|YX\rangle$ that form degenerate ground-states at $h^\star_{xy}$ cease to be exact ground-states, although the ground-state in the soliton phase is always two-fold degenerate. Instead, an approximation to the ground-states can be found by considering the closely related product states of the form
\begin{equation}
    |X'Y'\rangle=|x'y'x'y'\ldots\rangle,\ \ |Y'X'\rangle=|y'x'y'x'\ldots\rangle.
\end{equation}
where the states $|x'\rangle$ and $|y'\rangle$ are not orthogonal but instead at an angle {\it exceeding} 90 degrees by a small amount $\delta$ in either direction, justifying the $|x'\rangle$, $|y'\rangle$ notation.
For such states
the spins are partly aligned with the field and the Zeeman term can still lower the energy considerably, however, as long as
$h_{xy}<h^\star_{xy}$ an additional lowering of the energy can be obtained from the Kitaev term if $\delta>0$.
If we consider a small $\delta>0$ then to linear order, each Kitaev term then lowers the energy by $-K S^2\delta$ while the average Zeeman term will change to $-Sh_{xy}(1-\delta)/\sqrt{2}$ increasing the energy by $+Sh_{xy}\delta/\sqrt{2}$. Hence, if
$h_{xy}<h^\star_{xy}$, a non-zero $\delta>0$ can lower the energy justifying the notation $|x'\rangle$ and $|y'\rangle$. 
For $h_{xy}>h^\star_{xy}$, $\delta$ changes sign and the angle between $|x'\rangle$, $|y'\rangle$ is smaller than 90 degrees quickly approaching the PS state which is reached when $\delta=-\pi/4$.
We note that,
for small $\delta$, the states $|X'Y'\rangle$ and $|Y'X'\rangle$ are still degenerate and linearly independent but no-longer orthogonal.

The presence of a non-zero $\delta$ implies that the soliton mass, $\Delta_b$ and anti-soliton mass, $\Delta_B$ vary with $h_{xy}$, as does the energy of the states $|Y'X'\rangle$ and $|X'Y'\rangle$ with respect to which they are defined. As we discuss further in section~\ref{sec:variational} it is possible to perform variational calculations to determine the optimal $\Psi_b$ and $\Psi_B$ as a function of $h_{xy}$ thereby obtaining variational estimates for the masses $\Delta_b^{var}$ and $\Delta_B^{var}$ versus $h_{xy}$. Such estimates should be relatively precise, close to $h^\star_{xy}$ progressively failing as the field is tuned away from $h^\star_{xy}$. If we use Eq.~(\ref{eq:soldef}) to define the soliton phase we can then use $\Delta_b^{var}$+$\Delta_B^{var}$$>0$ to estimate the extent of the soliton phase. Our variational results (see section~\ref{sec:variational}) for $\Delta_B^{var}$ and $\Delta_B^{var}$ are shown in Fig.~\ref{fig:VarSpinGap} for \sone\ as a function of $h_{xy}$ along
with their sum. Crucially, there is only a finite range around $h^\star_{xy}$ where $\Delta_b^{var}$+$\Delta_B^{var}$$>0$ and the soliton phase is stable, indicating a lower, $h_{xy}^{c1}$,  and upper $h_{xy}^{c2}$ critical field. The critical fields can also be determined very precisely from iDMRG calculations, which are indicated as the dotted red lines in Fig.~\ref{fig:VarSpinGap}. The variational estimate for
$h_{xy}^{c2}$ is in surprisingly good agreement with the iDMRG result, while the variational estimate of $h_{xy}^{c1}$ is significantly worse.
As we discuss in section~\ref{sec:results}, the agreement of the variational estimates with precise DMRG results for $\Delta_b$ progressively worsens as the field is tuned away from $h^\star_{xy}$. 
Nevertheless, the fact that the simple variational calculations predict the existence of a non-zero lower critical field, $h_{xy}^{c1}$, is highly non-trivial and consistent with the fact that the soliton phase appears as an island in the polarized sea (the PS state).

\subsection{The Kitaev Chain at $\mathbf{h}$=0}\label{sec:h0}
\begin{figure}
  \includegraphics[width=\columnwidth]{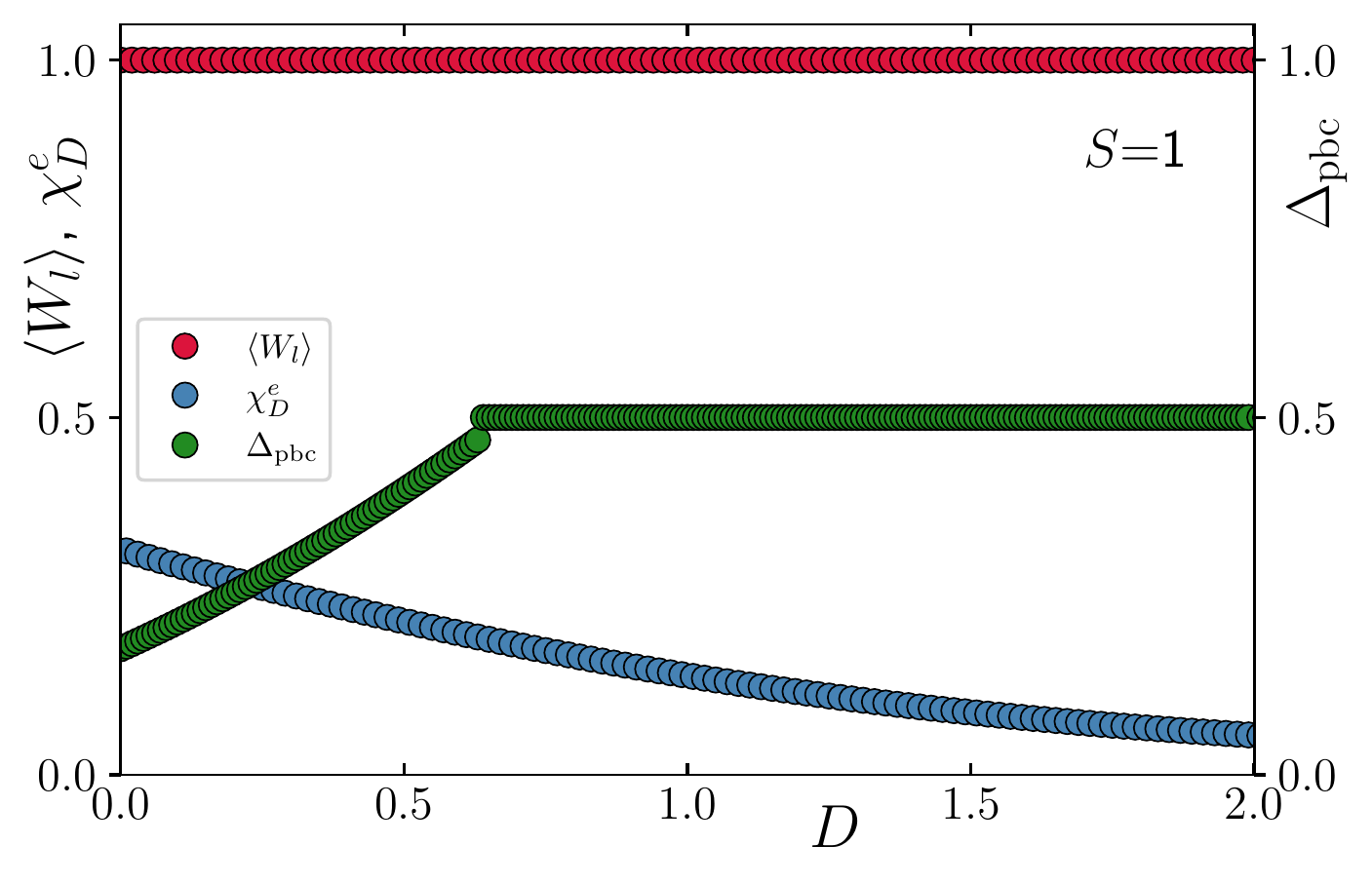}
  \caption{
  iDMRG results for the bond-parity operator $\langle W_l\rangle$ and susceptibility $\chi^e_D$
as a function of $D$ for the \sone\ Hamiltonian $\H_D$, Eq.~(\ref{eq:HD}). Results are shown alongside finite DMRG results with PBC for the spin gap, $\Delta_\mathrm{pbc}$ for $N$=60. A smooth evolution with $h_{xy}$ is evident and no transition is observed.
  }
  \label{fig:WllargeD}
\end{figure} 
The Kitaev chain in zero field has a number of invariants similar to the plaquette operators defined for the Honeycomb model~\cite{kitaev2006}.
As shown in Ref.~\onlinecite{Sen2010}, if site operators
\begin{equation}
    \mathcal{R}^x_l=e^{i\pi S^x_l}\ \ \ ,\ \ \ \mathcal{R}^y_l=e^{i\pi S^y_l}
\end{equation}
are defined, then, with $x$-bond ($y$-bond) couplings in $\H$, Eq.~(\ref{eq:H}), between $[l,l+1]$ with $l$ odd(even), bond-parity operators $W_l$ can be defined on odd and even bonds\cite{You2020}
\begin{equation}
   W_{2l-1}=\mathcal{R}^y_{2l-1} \mathcal{R}^y_{2l},\ \ W_{2l-1}=\mathcal{R}^x_{2l} \mathcal{R}^x_{2l+1},\label{eq:bondop}
\end{equation}
that commutes with the Hamiltonian, $[W_l,\H]$=0, and for integer $S$, amongst themselves $[W_l,W_k]$=0. The $W_l$ are therefore invariants
and it can be shown that the ground-state lies in the sector with all $\langle W_l\rangle$=1 and for PBC it is non-degenerate. For half-integer $S$, $W_l$ anti-commutes with, $W_{l\pm 1}$ making  the physics of the half-integer spin Kitaev chain distinct from the case of integer $S$ that we consider here.

In materials other interactions than the Kitaev interactions will be present and the \sone\ Kitaev chain has been studied in the presence
of an additional Heisenberg coupling, $J$~\cite{You2020,You2022}, a $\Gamma$-term~\cite{Luo2021} and also in the presence of anisotropy~\cite{Liu2015,Gordon2022}. However, it is important to consider in detail the nature of the zero field ground-state of the isotropic \sone\ chain with $J$=$\Gamma$=0. In Ref.~\cite{You2020,You2022} the ground-state at $\mathbf{h}$=0 was described as a quantum spin liquid, however, in Ref.~\cite{Luo2021} it was noted that the entanglement spectrum
is not doubled and concluded it is not a symmetry protected topological (SPT) state~\cite{Schuch2011,Chen2011,WenRMP2017}. 
Following Ref.~\onlinecite{Pollmann2012b} we have therefore investigated the projective representations, $U$ that can be obtained from the mixed transfer matrices in iDMRG. In general, if the site symmetries, $\mathcal{R}^x$ and $\mathcal{R}^y$ are respected
their representations can differ by a phase that must be $\pm 1$:
\begin{equation}
U(\mathcal{R}^x)U(\mathcal{R}^y)=\pm U(\mathcal{R}^y)U(\mathcal{R}^x).
\end{equation}
It is then convenient to isolate the phase factor by defining~\cite{Pollmann2012b}:
\begin{equation}
\mathcal{O}_\mathrm{Z_2\times Z_2}\equiv\frac{1}{\chi}\Tr\left( U(\mathcal{R}^x)U(\mathcal{R}^y) U^\dagger (\mathcal{R}^x)U^\dagger (\mathcal{R}^y)\right),
\end{equation}
with $\chi$ the bond dimension.
For the \sone\ Kitaev chain at $\mathbf{h}$=0 we find $\mathcal{O}_\mathrm{Z_2\times Z_2}=1$.
Similarly, under time reversal one finds that at $\mathbf{h}$=0 
\begin{equation}
\mathcal{O}_\mathrm{TR}\equiv\frac{1}{\chi}\Tr\left( U_\mathrm{TR}U_\mathrm{TR}^\star \right)=1,
\end{equation}
with $\star$ denoting complex conjugation and $\chi$ the bond dimension. Finally, if inversion is considered, one again finds that the trivial phase factor $\mathcal{O}_\mathcal{I}$=1. This is in contrast to the Haldane phase of the \sone\ spin chain where it is known that $\mathcal{O}_\mathrm{Z_2\times Z_2}$=$-1$, $\mathcal{O}_\mathrm{TR}$=$-1$
in addition to a non-trivial phase factor of $\mathcal{O}_\mathcal{I}$=$-1$ when considering inversion~\cite{Pollmann2010,Pollmann2012a}.
For \sone\ we can illustrate the trivial nature of the ground-state of the Kitaev chain at $\mathbf{h}$=0 by adding an easy-plane crystal field term, $D$ of the form $D\sum_j (S^z_j)^2$ to the $\mathbf{h}$=0 Hamiltonian to 
obtain 
\begin{equation}\label{eq:HD}
\H_D = K\sum_{j}\left(S_{2j+1}^xS_{2j+2}^x + S_{2j+2}^yS_{2j+3}^y\right)+ D\sum_j (S^z_j)^2.
\end{equation}
Note that, the $D$-term preserves the symmetries present at $\mathbf{h}$=0 in Eq.~(\ref{eq:H}). 
In the $D\to\infty$ limit, the ground-state
of Eq.~(\ref{eq:HD}) is the trivial product-state $|0\rangle|0\rangle|0\rangle\ldots$. We can now study the evolution of $\H_D$ as $D$ is increased from zero. In Fig.~\ref{fig:WllargeD} we show iDMRG results for $\langle W_l\rangle$ which remain a constant $\langle W_l\rangle$=1 for any  $D$.
The gap $\Delta_\mathrm{pbc}$ increases with $D$ and never approaches zero, likewise, the energy susceptibility $\chi^e_D$ quickly goes monotonically to zero. The evolution is smooth, and no transition is observed, consistent with the trivial nature of the ground-state at $\mathbf{h}$=0. Without breaking the symmetry, we have connected the two states. This defines what is sometimes called a symmetry protected trivial phase~\cite{Fuji2015,Kshetrimayum2016} (SPt) or alternatively a trivial SPT phase~\cite{Tsui2015}.
\begin{figure}
  \includegraphics[width=\columnwidth]{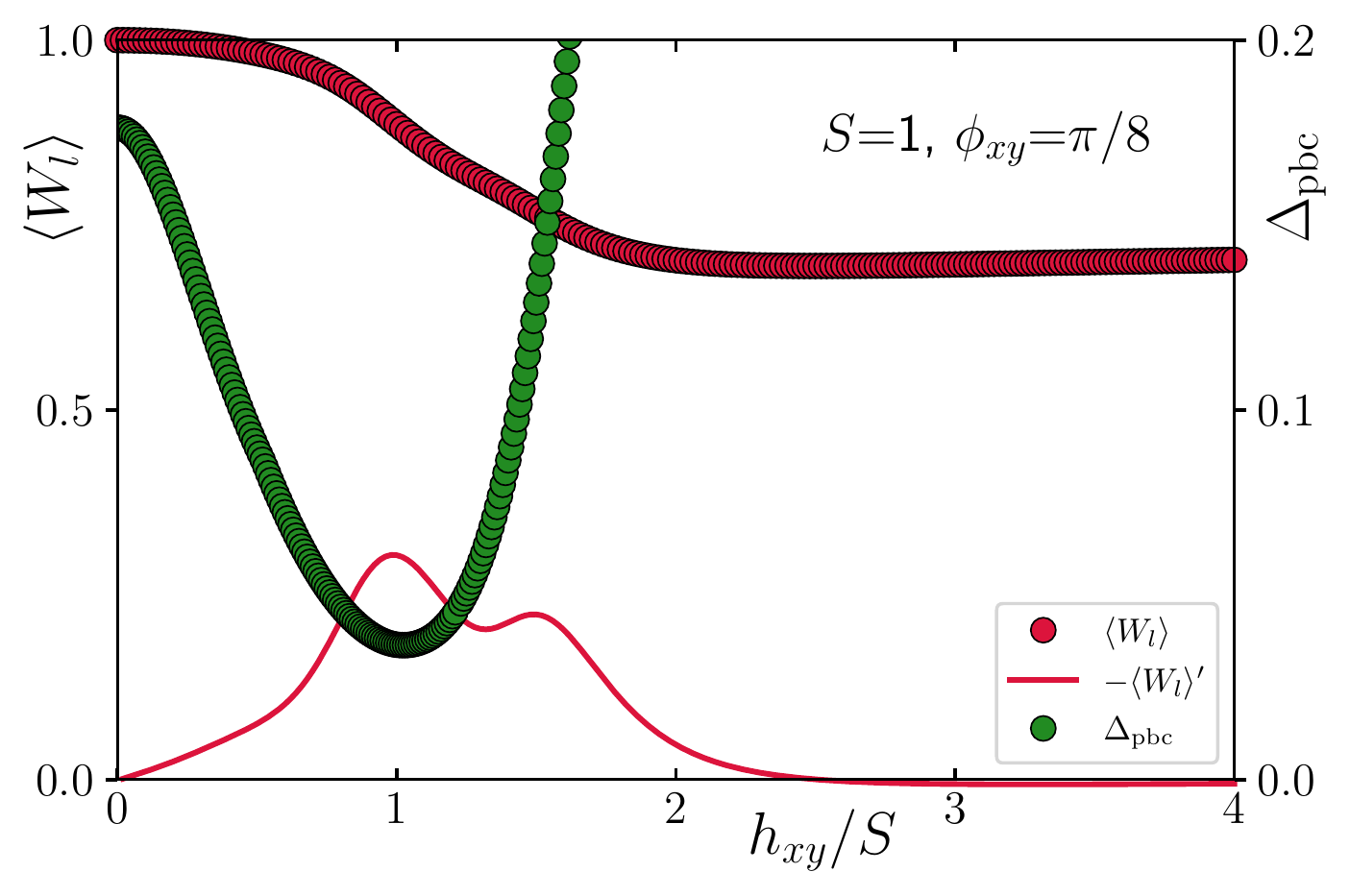}
  \caption{
  iDMRG results with \sone\ for the bond-parity operator $\langle W_l\rangle$ 
 and its derivate $\langle W_l\rangle^\prime$
as a function of $h_{xy}/S$ at an angle $\phi_{xy}=\pi/8$ in the $h_x, h_y$ plane shown alongside finite DMRG results with PBC for the spin gap, $\Delta_\mathrm{pbc}$ for $N$=60. A smooth evolution with $h_{xy}$ is evident and no transition is observed.
  }
  \label{fig:Wl}
\end{figure}

It is known that any SPT phase can be connected to the same trivial product state if we break the symmetry~\cite{Verstraete2004,Chen2011,Schuch2011,Chen2013}. 
In our determination of the phase diagram in section~\ref{sec:phasediag} this turns out to be an important point since, as already shown in Fig.~\ref{fig:phasediagram}, the soliton phases appear as isolated islands within the polarized state implying that a path can be found between the $\mathbf{h}$=0 and $h_{xy}$=$\infty$ ground-states without an intervening phase transition.
We note that, in contrast to the $D$ term discussed above, the introduction of a field term at a general angle will break most symmetries present in the Hamiltonian, Eq.~(\ref{eq:H}).
For \sone\ we can demonstrate the absence of a transition by calculating $\langle W_l\rangle$ and $\Delta_{pbc}$ as a function of $h_{xy}$ 
which should interpolate smoothly between $\mathbf{h}$=0 and the large field limit where the simple product state associated with complete field polarization is the ground-state. 
iDMRG results for such a calculation are shown in Fig.~\ref{fig:Wl} where $\langle W_l\rangle$ 
is graphed versus $h_{xy}/S$ along with finite DMRG results for the spin gap, $\Delta_\mathrm{pbc}$ for $N$=60. The calculations are done at a fixed angle $\phi_{xy}$=$\pi/8$ shown as the dotted blue line in Fig.~\ref{fig:phasediagram}, that does not intersect with the soliton phase for \sone. As is clear from the results in Fig~\ref{fig:Wl} the evolution is smooth, and no transition is observed, although some structure in $\langle W_l\rangle^\prime$ can be observed in the proximity of the soliton phase where $\Delta_\mathrm{pbc}$ also has a minimum.
In summary, for \sone\ we therefore conclude that the $\mathbf{h}$=0 phase is a symmetry protected trivial (SPt) phase.
Once the field is applied in a general direction, the symmetry is broken, and there is no distinction between the SPt and polarized states. 
However, along the unique directions $h_x=\pm h_y$ a transition to the soliton phase is possible since the chain is still protected by
the combined symmetry operation of a rotation on each site by $\pi$ around the field direction, ${\cal R}^{xy}$=$\exp(i\pi (S^x+S^y)/\sqrt{2})$, followed by a translation by one lattice spacing, $T$.
We expect this to hold for all integer $S$ but the half-integer case is distinct,
as discussed in~\cite{Sorensen2022} for \shalf, since the  ${\cal R}^{xy}\otimes T$ symmetry protection allow for a critical line to be present along the $h_x=\pm h_y$ symmetry directions, connecting the soliton phase to $\mathbf{h}$=0.

\section{Numerical Methods}\label{sec:num}
In the following we present results mainly obtained from 
finite size density matrix renormalization group~\cite{White1992a,White1992b,White1993,Schollwock2005,Hallberg2006,Schollwock2011} (DMRG) using both periodic (PBC) and open (OBC) boundary conditions as well as from infinite DMRG~\cite{McCulloch2008,Schollwock2011} (iDMRG) techniques. 
For the iDMRG calculations, we use a unit cell of either 12 or 24 sites.
We note that well converged iDMRG results should yield results in the thermodynamic limit free of finite-size effects independent of the size of the unit cell.
Typical precisions for both DMRG and iDMRG are $\epsilon<10^{-11}$ with a bond dimension in excess of 1000.
In order to establish the phase diagram, we focus on the following susceptibilities. With $e_0$ the ground-state energy per spin, we define the energy susceptibilities
\begin{equation}
  \chi_h^e = -\frac{\partial^2 e_0}{\partial h^2},\ \ \chi_{\phi_{xy}}^e = -\frac{\partial^2 e_0}{\partial \phi_{xy}^2}, \ \ 
  \chi_{\theta_z}^e = -\frac{\partial^2 e_0}{\partial \theta_z^2}
\end{equation}
where $h$ is the field strength and $\phi_{xy}$ and $\theta_z$ the field angles.
Here, $\chi_h^e$ is effectively a magnetic susceptibility.
At a quantum critical point (QCP) it is known~\cite{Albuquerque2010} that,
for a finite system of size $N$, the energy susceptibility diverges as
\begin{equation}
  \chi^e \sim N^{2/\nu-d-z}.
\end{equation}
Here $\nu$ and $z$ are the correlation and dynamical critical exponents and $d$ is the dimension.
We see that 
$\chi^e$ only diverges at the phase transition if the critical exponent $\nu$ is smaller than $2/(d+z)$. In the present
case $d$=$1$ and we assume $z$=1, so $\nu<1$ if a divergence is observed.

In section~\ref{sec:specheat} we present thermodynamic results for the specific heat  as a function of temperature.
The results are obtained using purification~\cite{Uhlmann1976,Uhlmann1986,Verstraete2004,Barthel2009,Karrasch2012,Barthel2017,Hauschild2018} where
the density matrix  $\rho$ acting on a physical Hilbert space $\H^P$ is represented as a pure state $|\psi\rangle$ in an enlarged space $\H^P\otimes\H^A:$
\begin{equation}
\rho = \Tr_A|\psi\rangle\langle\psi|,
\end{equation}
where the ancillary space $\H^A$ can be taken to be identical to $\H^P$.
This gives the 
thermofield double purification~\cite{Takahashi1975,Israel1976} (TFD)
\begin{equation}
|\psi_\beta\rangle = \frac{1}{\sqrt{Z}}\sum_ne^{-\beta E_n/2}|n\rangle_P|n\rangle_A,
\end{equation}
where $|n\rangle$ are the eigenvectors and $E_n$ the eigenvalues of $\H$ and thermal expectation values of an operator $\mathcal{O}$ can be obtained from $\langle\psi_\beta|\mathcal{O}|\psi_\beta\rangle$. The TFD can be obtained by using imaginary-time evolution 
$|\psi_\beta\rangle\sim e^{-\beta \H/2}|\psi_0\rangle$ 
starting from a state $|\psi_0\rangle$=$\prod_i\frac{1}{\sqrt{d}}\sum_\mathbf{\sigma_i}|\mathbf{\sigma_i}\rangle_P|\mathbf{\sigma_i}\rangle_A$, where $\mathbf{\sigma_i}$ runs over the local Hilbert space of dimension $d$.  On a given site, the physical and ancillary degrees of freedom are then maximally entangled in the state $|\psi_0\rangle$. For the calculations presented in section~\ref{sec:specheat} imaginary time evolution with a time step of $0.001$ is used.

\section{DMRG and iDMRG Results}\label{sec:results}
\subsection{Phase Diagram}\label{sec:phasediag}
\begin{figure}
  \includegraphics[width=\columnwidth]{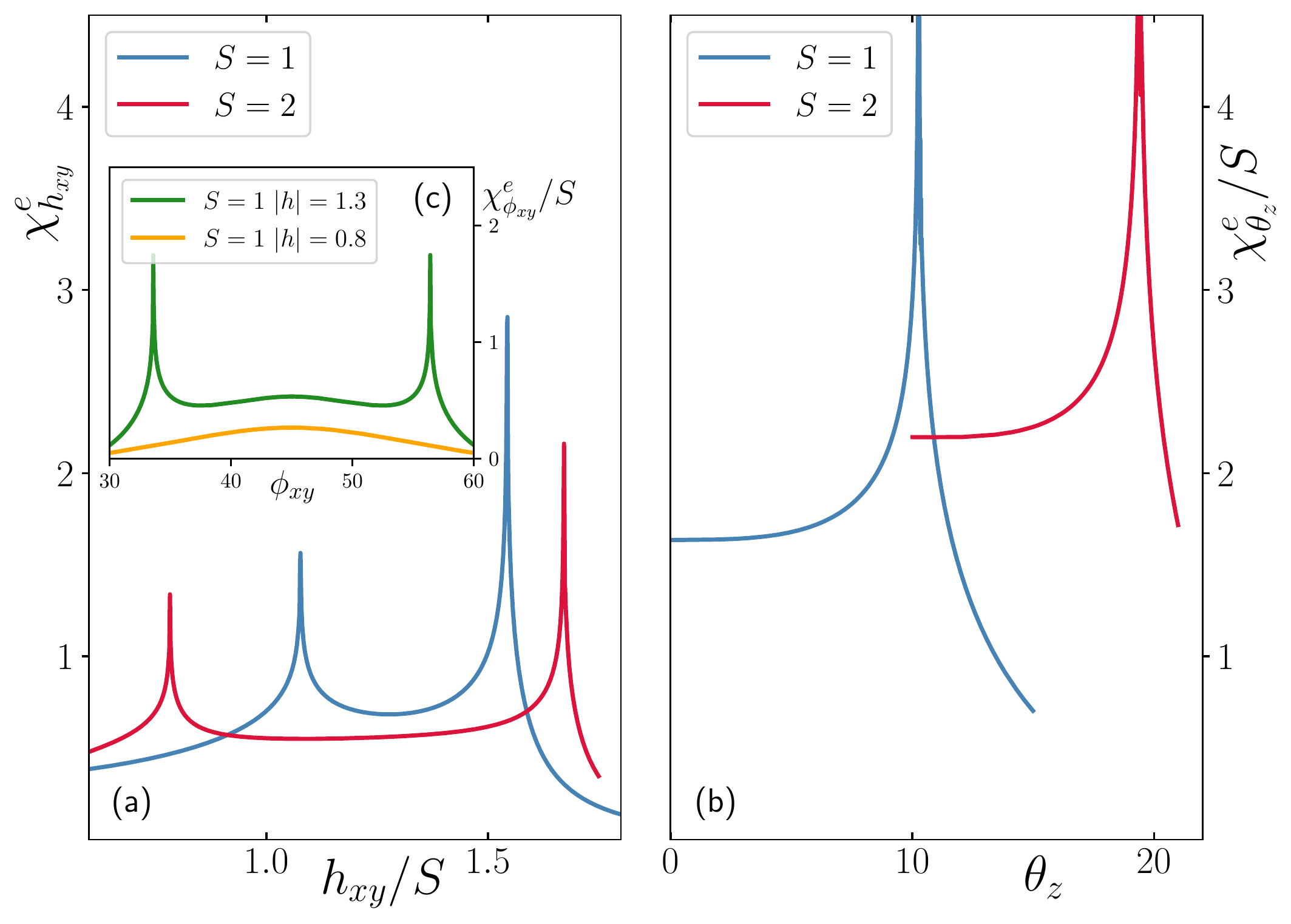}
  \caption{(a) iDMRG results for $\chi^e_{h_{xy}}$  versus the field strength $h_{xy}/S$ for the \sone\  and \stwo\ Kitaev spin chains, showing the positions of the
  critical fields $h^{c1}_{xy}$ and $h^{c2}_{xy}$.
  (b) iDMRG results for the \sone\  and \stwo\ Kitaev spin chains, showing
    $\chi^e_{\theta_{z}}/S$ versus the field angle $\theta_{z}$ for field strengths of $|h|/K$=$1.3$ (\sone\ ) and $2.6$ (\stwo\ ).
  (c) iDMRG results for the \sone\ Kitaev spin chain for
    $\chi^e_{\phi_{xy}}$ versus the field angle $\phi_{xy}$ for field strengths of $|h|/K$=$0.8$ and $1.3$. Note the absence of transitions for $|h|/K=0.8$
  }
  \label{fig:chie}
\end{figure}  
Our results for the phase diagram for \sone, \stwo, $S$=3, and to a lesser extent also for $S$=4,5 are summarized in Fig.~{\ref{fig:phasediagram}} where the extent of the soliton phase in the $h_x, h_y$ plane is shown as obtained from iDMRG results for $\chi^e_h$ and $\chi^e_{\phi_{xy}}$. 
Remarkably, the soliton phase appears as an {\it island} in the polarized sea since the PS state completely surrounds the soliton phase, as we have discussed above. For \sone\ this is illustrated in Fig.~\ref{fig:chie}(c) where $\chi^e_{\phi_{xy}}$ is shown for the field values $h_{xy}/K$=0.8 and 1.3. As the field angle $\phi_{xy}$ is varied, clear transitions are visible  for $h_{xy}/K$=1.3, but completely absent for $h_{xy}/K$=0.8. If instead the field strength, $h_{xy}$
is varied at a field angle of $\phi_{xy}$=$\pi/4$ then two very well-defined transitions are clearly visible in Fig.~\ref{fig:chie}(a) for both \sone\ and \stwo. The peak positions are what is plotted in Fig.~\ref{fig:phasediagram}. We have extensively search for a phase transition distinguishing the low-field phase ($h_{xy}<h^{c1}_{xy}$) from the PS state using different techniques and different paths through the phase diagram, but it appears adiabatically connected to the PS phase as explicitly shown in section~\ref{sec:h0}.
This is likely unique to the integer spin models, since for $S$=1/2 results indicate the presence of a critical line~\cite{Sorensen2022} for $h_{xy}<h^{c1}_{xy}$.

The soliton phase is not only restricted to the $h_x,h_y$ plane, but extends to non-zero $\theta_z$.
This is demonstrated in Fig.~\ref{fig:chie}(b) where iDMRG results for  $\chi^e_{\theta_{z}}$ versus $\theta_z$ are shown at the fixed field values of $h_{xy}/K$=$1.3$ and $h_{xy}/K$=$2.6$ for \sone\ and \stwo\ respectively. Clear transitions are observed at the critical angles $\theta_z$=$10.27^\circ$ (\sone\ ) and $19.41^\circ$ (\stwo\ ).

\subsection{Energy Gaps}\label{sec:gaps}
\begin{figure}
    \includegraphics[width=\columnwidth]{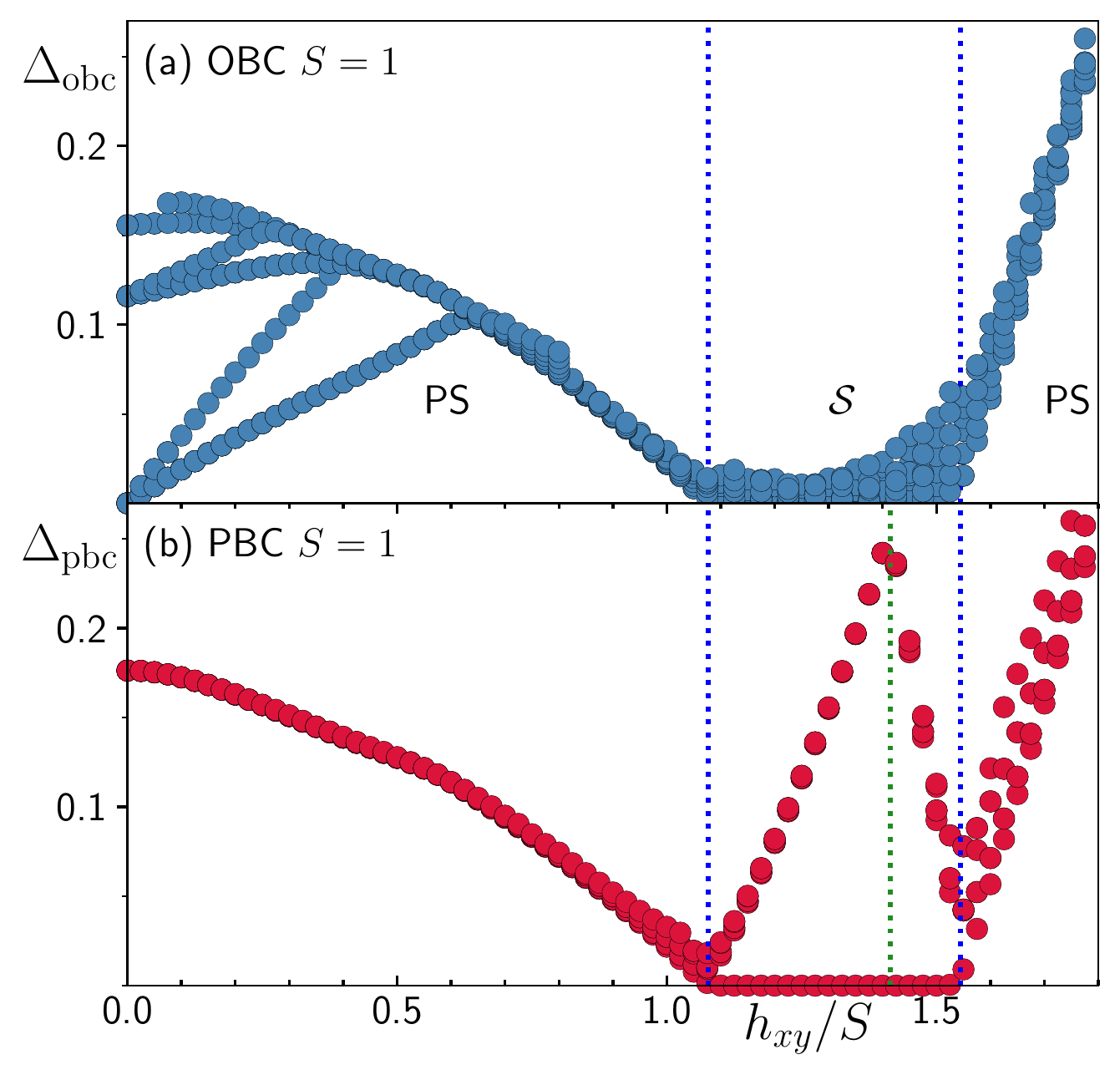}
    \caption{DMRG results for the first few excited states as a function of field, $h_{xy}$ at $\phi_{xy}$=$45^\circ$ and $\theta_z$=$0$ for the \sone\ Kitaev spin chain. The critical fields delineating the soliton phase are indicated by the dotted blue lines.
    (a) Results for OBC with $N$=$100$. At $h_{xy}$=$0$ the ground-state is four-fold degenerate. Note the proliferation of low-lying states in the soliton phase, marked by '$\mathcal{S}$'.
    (b) Results for PBC with $N$=$60$. Note, the two-fold degenerate ground-state in the soliton phase. The green dotted line indicates $h^\star_{xy}$=$SK\sqrt{2}$.
    }  
    \label{fig:gaps}
\end{figure}
We next turn to a discussion of the energy spectrum at fixed field angles $\theta_z$=$0$, $\phi_{xy}$=$\pi/4$ as a function of field strength, $h_{xy}$ and for brevity we only discuss the \sone\ chain. Due to the rapid growth of the size of the Hilbert space with $N$ it is convenient to use finite size DMRG calculations to determine the ground- ($E_0$) and excited- ($E_n$) state energies and study the gaps ($\Delta_n$=$E_n$-$E_0$) in the spectrum. Our results are shown in Fig.~\ref{fig:gaps}.

We first focus on PBC, where our results are shown in Fig.~\ref{fig:gaps}(b). We exclusively consider, $N$ even dictated by the two-site unit cell.
The ground-state at $h_{xy}$=$0$ is non-degenerate below a sizable gap, $\Delta_\mathrm{pbc}(h_{xy}$=$0)$=$0.1763 K$ in agreement with previous results~\cite{You2020}. The first excited state at $h_{xy}$=$0$ is known to be $N$-fold degenerate~\cite{Sen2010,You2020}. At $h^{c1}_{xy}$=$1.077K$ the gap closes, and the soliton phase is entered.
Within the soliton phase for $h^{c1}_{xy}<h_{xy}<h^{c2}_{xy}$=$1.544K$ the ground-state is exactly two-fold degenerate, even for finite $N$, below a sizable gap. As mentioned previously, the maximum of the gap coincides with the presence of the two exact product ground-states $|YX\rangle$ and $|XY\rangle$ at $h^\star_{xy}$ (indicated as the green dotted line in  Fig.~\ref{fig:gaps}(b)) where the gap is estimated to be $\Delta_\mathrm{pbc}(h_{xy}$=$0)$=$0.2555 K$.

We then turn the attention to  OBC (Fig.~\ref{fig:gaps}(a)) where the ground-state at $\mathbf{h}$=$0$ is four-fold degenerate for \sone\cite{You2020,Gordon2022}. For small fields, the ground-state degeneracy is lifted, and a low-lying doublet appears below a singlet. At field strengths $h_{xy}\sim 0.4-0.6$ the low-lying singlet and doublet merge with the other low-lying states which we assume might form the lower edge of a continuum.
When the lower critical field $h^{c1}_{xy}$ is reached the gap closes and throughout the soliton phase, marked as $\mathcal{S}$ in Fig.~\ref{fig:gaps}, a proliferation of low-lying states is visible until the upper critical field $h^{c2}_{xy}$=$1.544K$ is reached where the chain transitions back into the polarized state and a gap opens up. Within the soliton phase the DMRG results for the gaps indicate significant finite-size corrections which we have not been able to analyze in detail, and it has not been possible to determine if these low-lying states correspond to a true gapless spectrum as opposed to a significant number of discrete in-gap levels appearing within the gap present for periodic boundary conditions. The ground-state degeneracy, if any, within the soliton phase for OBC is also an open question.

The difference in the spectrum within the soliton phase is rather remarkable, even more so since the spectrum for OBC does not depend on the parity of $N$ and occurs equally well for $N$ even and odd. As discussed in the introduction, the absence of SU(2) symmetry means that it is difficult to explain the multitude of low-lying states occurring for OBC as arising from unpaired degrees of freedom.

\begin{figure}
  \includegraphics[width=\columnwidth]{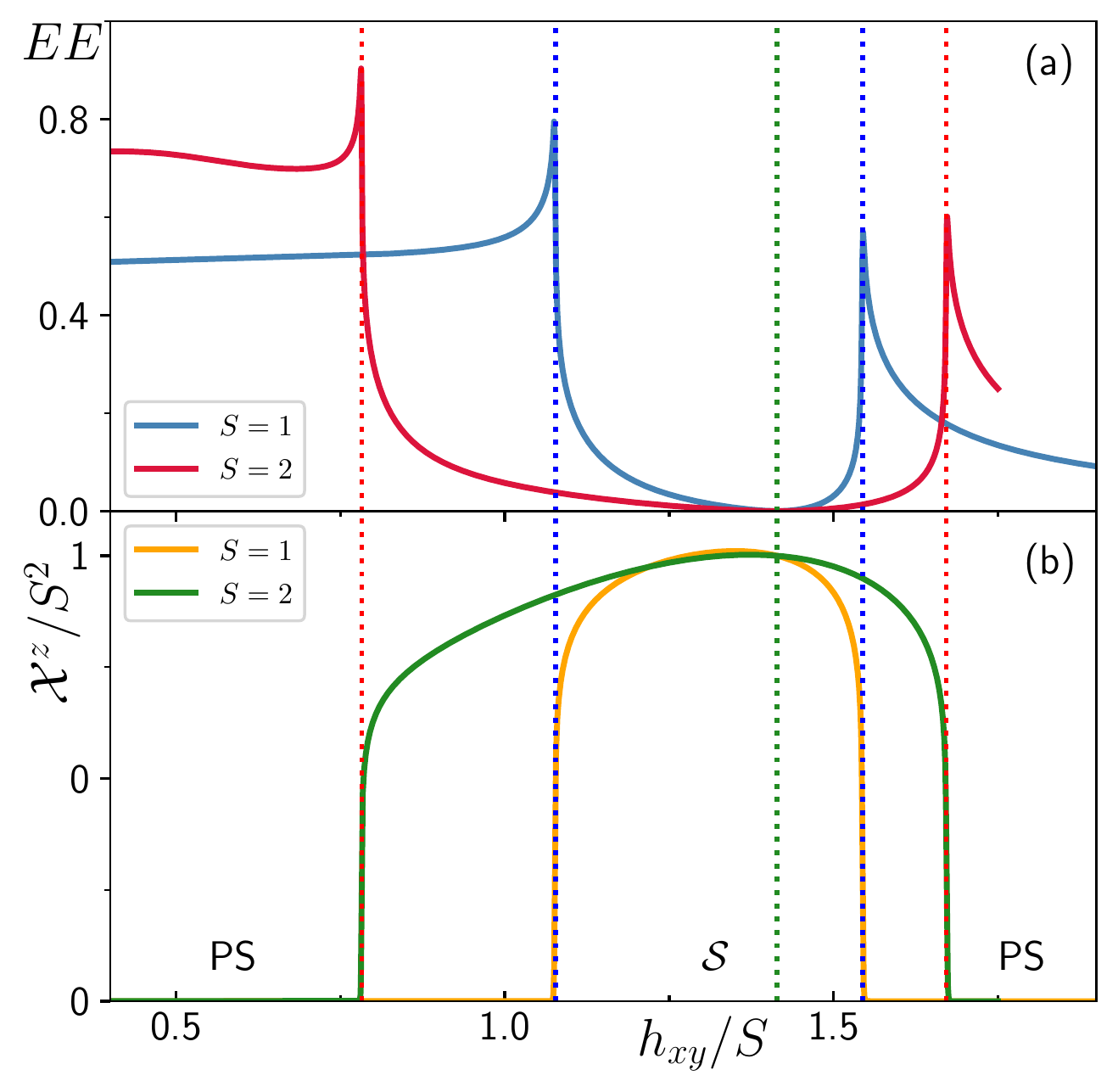}
  \caption{
  iDMRG results for the \sone\ and \stwo\ Kitaev spin chains. 
  The dashed lines indicate the critical fields $h^{c1}_{xy}$ and $h^{c2}_{xy}$. 
  (a) The entanglement entropy $EE$  versus the field strength $h_{xy}/S$. Note that very low $EE$ in the soliton phase. 
  (b) The z-component of the vector chirality scaled with $S^2$, 
    $\mathcal{X}^z/S^2$, versus $h_{xy}/S$ .
  }
  \label{fig:chiralZ}
\end{figure}  
\subsection{Chiral Order, $\mathcal{X}^z$ and Entanglement}
In light of the two exact ground-sates $|YX\rangle$ and $|XY\rangle$ occurring at $h^\star_{xy}$ for PBC it is not
surprising that the soliton phase can be characterized by a non-zero vector
chirality, $\mathcal{X}^\alpha$: 
\begin{equation}
    \mathcal{X}^\alpha= (-1)^j\langle ({\bf S}_j \times {\bf S}_{j+1})^\alpha\rangle.
\end{equation}
While ${\mathcal{X}}^{x,y}$=$0$ in the soliton phase, ${\mathcal{X}}^z$$\neq$0 as was previously established for \shalf. This is shown in Fig.~\ref{fig:chiralZ}(b) for \sone\ and \stwo\ where iDMRG results for ${\mathcal{X}}^z$ are plotted as a function of $h_{xy}/S$. As can be seen, ${\mathcal{X}}^z$ remains sizable throughout the soliton phase reaching a maximum close to (or at) $h^\star_{xy}$ before abruptly going to zero at $h^{c1}_{xy}$ and $h^{c2}_{xy}$. The soliton phase should then be regarded as a chiral soliton phase.

In Fig.~\ref{fig:chiralZ}(a) we show results for the bipartite entanglement entropy:
\begin{equation}
EE = -\Tr \rho_A \ln \rho_A
\end{equation}
where $\rho_A$ is the reduced density for half the system.
The states $|YX\rangle$ and $|XY\rangle$ are only {\it exact } ground-states for PBC and the iDMRG results shown in  Fig.~\ref{fig:chiralZ}(a) are obtained for OBC. Hence at $h^\star_{xy}$, shown as the green dotted line in Fig.~\ref{fig:chiralZ}, the entanglement entropy $EE$ is not strictly zero, as should be the case for an exact product state, but rather extremely small. 
As is clearly visible
in Fig.~\ref{fig:chiralZ}(a), $EE$ peaks at $h^{c1}_{xy}$ and $h^{c2}_{xy}$ but away from the quantum critical points it
remains rather small throughout the entire soliton phase, approaching zero at $h^\star_{xy}$, implying that the ground-state is close to a product state within the soliton phase.

\subsection{Soliton Mass, $\Delta_b$ and Width, $\xi_\mathcal{S}$}\label{sec:solmass}
\begin{figure}
  \includegraphics[width=\columnwidth]{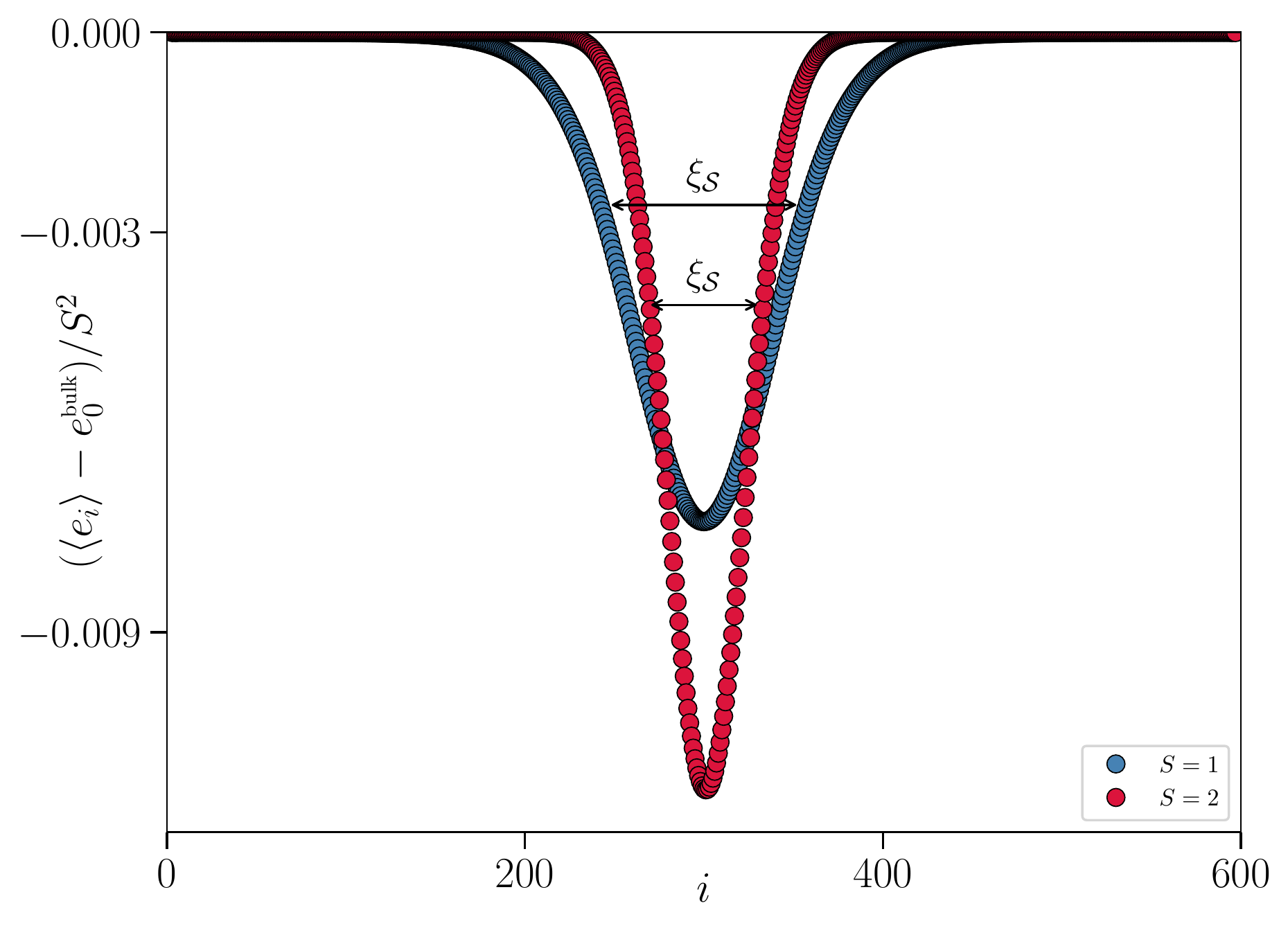}
  \caption{
   Finite size DMRG results with $N$=$600$ for the \sone\ (blue) and \stwo\ (red) Kitaev spin chains
   showing the relative energy density $(\langle e_i\rangle-e^{\mathrm{bulk}}_0)/S^2$ versus position, $i$,
   in the chain. Results are shown for $h_{xy}/K$=$1.32 (S=1)$ and $h_{xy}/K$=$2.60 (S=2)$
  }
  \label{fig:ei}
\end{figure}  
The variational calculation of the soliton mass for OBC described in section~\ref{sec:varsolmass} and \ref{sec:variational} relies on a subtractive procedure where the energy of the single soliton state is measured with respect to the isotropic product state. For a more detailed understanding of the DMRG results it is useful to have a more refined measure of $\Delta_b$ that does not involve a subtraction. In the absence of SU(2) symmetry and a well defined spin for the soliton it is then necessary to focus on the local bond energy density which we define as the energy of the bond $[i,i+1]$ plus $1/2$ the field terms on the sites $i$ and $i+1$. Far away from the soliton the energy density attains a constant value $e_0^{\mathrm{bulk}}$ and we expect that this bulk energy density is essentially identical to the energy density of the two fold degenerate ground-states with PBC. It is then instructive to study the following quantity:
\begin{equation}
\langle e_i\rangle-e_0^{\mathrm{bulk} }
\end{equation}
This is shown in Fig.~\ref{fig:ei} where $\langle e_i\rangle-e_0^{\mathrm{bulk}}$ is plotted versus $i$ for $h_{xy}/K$=$1.32$ (\sone)
and 2.60 (\stwo), showing a sharply localized soliton. Furthermore, the soliton 'sharpens' with increasing $S$, displaying a smaller spatial extent.
We can now simply define the soliton mass, $\Delta_b$, as the integrated deviation from $e_0^{\mathrm{bulk}}$ in the following manner:
\begin{equation}
\Delta_b=\sum_i \left( \langle e_i\rangle-e_0^{\mathrm{bulk}}\right).\label{eq:sumei}
\end{equation}
Clearly, this measures by how much the soliton has lowered the total energy which was our original definition of the soliton mass, $\Delta_b$.

\begin{figure}
  \includegraphics[width=\columnwidth]{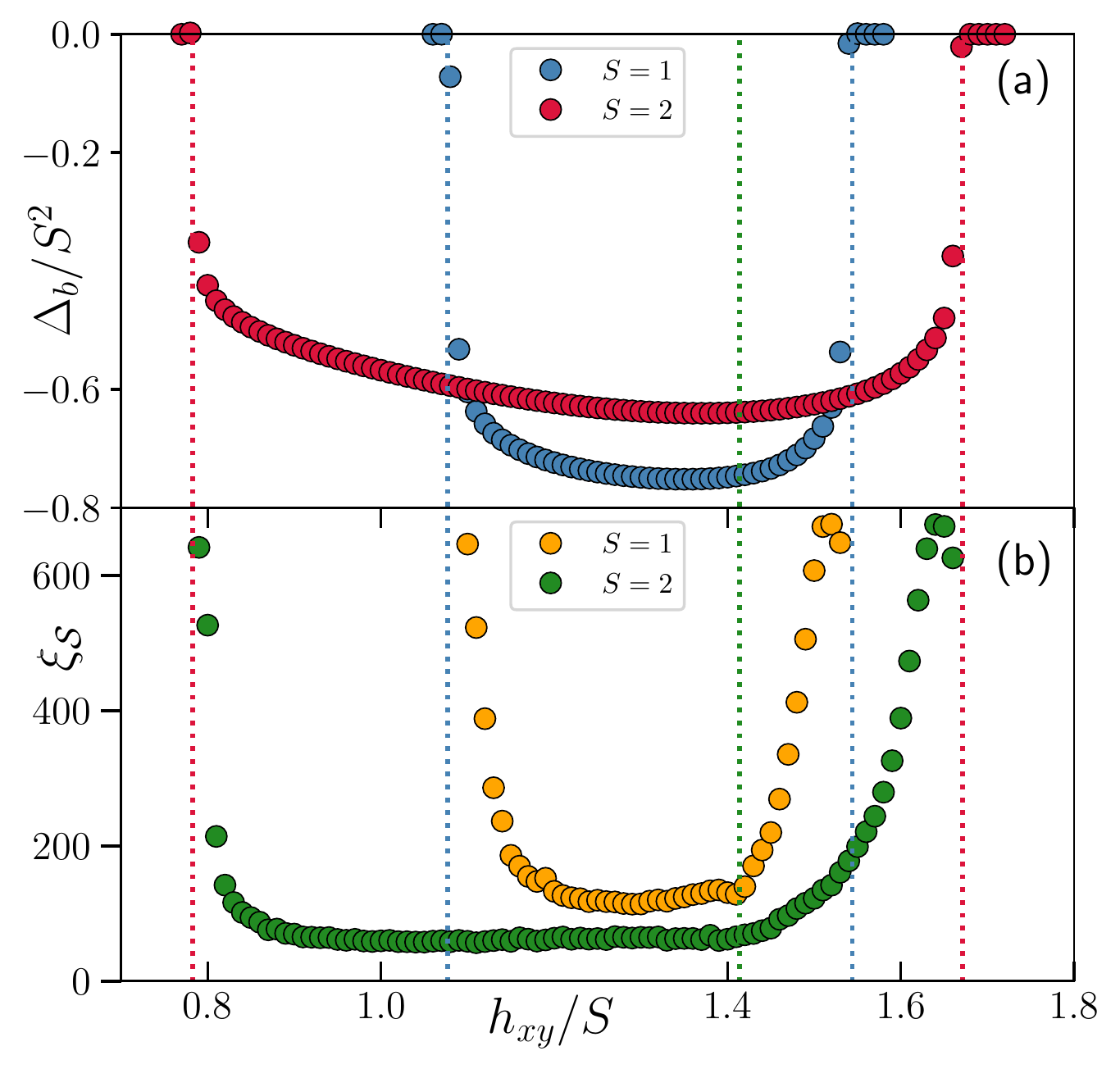}
  \caption{
   Finite size DMRG results with $N$=$1200$ for the \sone\ and \stwo\ Kitaev spin chains. 
  The dashed lines indicate the critical fields $h^{c1}_{xy}$ and $h^{c2}_{xy}$. 
  (a) The soliton mass, $\Delta_b/S^2$ $EE$  versus the field strength $h_{xy}/S$. 
  (b) The soliton size $\xi_{\scriptstyle \mathcal{S}}$ versus $h_{xy}/S$.
  }
  \label{fig:solitonmass}
\end{figure}  
From high precision DMRG calculations with OBC on $N$$=$1200 sites for a range of $h_{xy}$ we can now extract $\Delta_b$ for both \sone\ and \stwo.
Our results are illustrated in Fig.~\ref{fig:solitonmass}(a) where $\Delta_b/S^2$ is shown as a function of $h_{xy}$. As one might expect,
$\Delta_b$ is roughly proportional to $S^2$, consistent with classical models of solitons~\cite{Mikeska1991}, and with only a modest variation throughout the soliton phase. In contrast to the variational
results for $\Delta_b^\mathrm{var}$ shown in Fig.~\ref{fig:VarSpinGap} the DMRG results in Fig.~\ref{fig:solitonmass}(a) show that
$\Delta_b$ tends to zero at $h^{c1}_{xy}$ and $h^{c2}_{xy}$. From the definition, Eq.~(\ref{eq:sumei}) it follows that $\Delta_b$=$0$ outside
the soliton phase where we expect the energy density to be uniform. In contrast, the variational states $\psi_b$ can never yield a uniform
energy density, and we have to use a less refined measure for the soliton mass. However, it is still useful to compare the estimates at
$h^\star_{xy}$, where we from DMRG for \sone\ find $\Delta_b$$=$$-0.7457$ and from the variational calculations $\Delta_b^\mathrm{var}$=$-0.7225$, in good agreement.

The energy profiles shown in Fig.~\ref{fig:ei} can be used to estimate the size of the soliton, $\xi_\mathcal{S}$ by simply measuring at
what distance $|\langle e_i\rangle-e_0^{\mathrm{bulk}}|$ has decreased by a factor of $1/e$ from the maximum. Measures of $\xi_\mathcal{S}$
are indicated on Fig.~\ref{fig:ei}. Using this definition of $\xi_\mathcal{S}$ we have determined the size of the soliton throughout
the soliton phase from high precision DMRG calculations with OBC on $N$=1200 sites for both \sone\ and \stwo. The results are shown
in Fig.~\ref{fig:solitonmass}(b). Through most of the soliton phase $\xi_\mathcal{S}$ remains roughly constant at around 120 lattice spacings for \sone\ and approximately 60 lattice spacings for \stwo, before increasing dramatically close to $h^{c1}_{xy}$ and $h^{c2}_{xy}$.

\section{Uniform Product States}\label{sec:classical}
As already discussed in section~\ref{sec:phenom} the product states $|YX\rangle$ and $|XY\rangle$ play a crucial role in our understanding
of the soliton phase. For $\theta_z$=$0$, $\phi_{xy}$=$\pi/4$ at $h^\star_{xy}$ they are exact ground-states for PBC, however, as pointed out in
section~\ref{sec:hc}, when $h_{xy}$ is
tuned away from $h^\star_{xy}$ a good approximation to the ground-state can be obtained by considering product states of the form $|Y'X'\rangle$
and $|X'Y'\rangle$ where the angle between $|y'\rangle$ and $|x'\rangle$ deviates from $\pi/2$ in both directions by an amount $c$. We now wish to establish a reliable estimate of the optimal value for this angle, $c^\star$, as a function of $h_{xy}$ for any $S$.

\subsection{Estimate of $c^\star$}
In the following we focus on the case of \sone\ and \stwo\ with generalizations to $S>2$ straight forward.
With $c$ taking the place of $\delta$ discussed in section~\ref{sec:hc}, we define for \sone\ the following states on a given site:
\begin{eqnarray}
|x'\rangle &=&(e^{i2c},\sqrt{2}e^{ic},1)/2\nonumber\\
|y'\rangle &=&(e^{-i2b},\sqrt{2}e^{-ib},1)/2,
\end{eqnarray}
with $b=\pi/2+c$,
while for \stwo\ we define:
\begin{eqnarray}
|x'\rangle &=&(e^{i4c},2e^{i3c},\sqrt{6}e^{i2c},2e^{ic},1)/4\nonumber\\
|y'\rangle &=&(e^{-i4b},2e^{-i3b},\sqrt{6}e^{-i2b},2e^{-ib},1)/4.\nonumber\\
\end{eqnarray}
We can then define the product states:
\begin{equation}
    |X'Y'\rangle=|x'y'x'y'\ldots\rangle,\ \ |Y'X'\rangle=|y'x'y'x'\ldots\rangle,
\end{equation}
for both \sone\ and \stwo.
The optimal value for the excess angle, $c^\star$, will depend on the field $h_{xy}$. However, if we neglect boundary effects, then, due to the simple product nature of the states, it is only necessary to consider a two site system in order to find the optimal $c^\star$. To proceed, we focus on a $x$-bond and assign half a field term to each bond and write the single bond Hamiltonian as follows:
\begin{equation}
    \H_{\rm 1 bond}=KS^x_1S^x_2-h_{xy}(S^x_1+S^y_1+S^x_2+S^y_2)\frac{1}{2\sqrt{2}},
\end{equation}
with the $1/\sqrt{2}$ arising from the field angle $\phi_{xy}$=$\pi/4$. 
Evaluating $E_{\rm 1bond}$=$<Y'X'|H_{\rm 2site}|Y'X'>$ we find:
\begin{equation}
    E_{\rm 1bond}=-KS^2\cos{c}\sin{c}-\frac{Sh_{xy}}{\sqrt{2}}(\cos{c}-\sin{c})
\end{equation}
Minimizing $E_{\rm 1bond}$ with respect to $c$ at a given $h_{xy}$
yields the optimal $c$ as
\begin{equation}
c^*=\tan^{-1}\left[\frac{u + \sqrt{4S^2 - u^2}}{  -u + \sqrt{4S^2 -u^2}}\right]=\cos^{-1}{\frac{u}{2S}}-\frac{\pi}{4},\label{eq:copt} 
\end{equation}
where $u$=$h_{xy}/K$. It follows that $c^*$ becomes zero at $u$=$h_{xy}/K$=$S\sqrt{2}$, coinciding with $h^\star_{xy}$ as, has to be the case. Furthermore, at $h_{xy}/K$=$2S$ the optimal value for $c$ becomes $c^*$=$-\pi/4$ and the spins are then fully aligned with the field for any $h_{xy}>2SK$. This signals the transition to the PS state at the classical level and is shown as the red dashed line in Fig.~\ref{fig:phasediagram}. Using the optimal value of $c^*$ from Eq.~(\ref{eq:copt}) one finds for the energy:
\begin{equation}
    E_{\rm 1bond}=-\frac{1}{4}(2S^2+u^2), \ \ u\le 2S.
\end{equation}

\subsection{Estimate of the product state defect energy}
It is illustrative to also consider a single defect state, at $h^\star_{xy}$ where calculations with the states $|\psi_b(i)\rangle$ can be 
significantly simplified. At $h^\star_{xy}$ we may estimate the defect energy of the
state
\begin{eqnarray}
|d\rangle=|
    \tcbset{colback=blue!10!white}
    \tcboxmath[size=fbox,auto outer arc, arc=5pt]{
    y\xbond\ 
    \nearrow_i\ybond\ x
    }
    \rangle,
\end{eqnarray}
and compare it to the state $|yxy\rangle$ on just 2 bonds sites since the two states will have the same energy elsewhere.
That is, we consider the 2 bond Hamiltonian:
\begin{flalign}
&\H_{\rm 2 bond}=KS^x_1S^x_2+KS_2^yS_3^y\nonumber &\\
&    -\frac{h_{xy}}{\sqrt{2}}(\frac{1}{2}(S^x_1+S^y_1)+S^x_2+S^y_2+\frac{1}{2}(S^x_3+S^y_3)),&
\end{flalign}
again counting the field terms on the first and last site by a factor of 1/2.
At $h^\star_{xy}$, it is straight forward to evaluate $\langle yxy|\H_{\rm 2bond}|yxy\rangle$=$-2$ and $\langle d|\H_{\rm 2bond}|d\rangle$=-1-$\sqrt{2}$.
The energy of the defect state $|d\rangle$ is then 1-$\sqrt{2}$$\sim$-0.4142 lower in energy than the  $|yxy\rangle$ state. As discussed in section~\ref{sec:solmass}, at $h^\star_{xy}$ DMRG results for $\Delta_b$ yields $-0.7457$, considerably lower.
Moreover, if this analysis is extended to $h_{xy}$$\neq$$h^\star_{xy}$, and to include the 
$
    |\tcbset{colback=red!10!white}
    \tcboxmath[size=fbox,auto outer arc, arc=5pt]{
    x\xbond\ 
    \nearrow_i\ybond\ y
    }
    \rangle
$
state describing the anti-defect, then the upper critical field coincides with the classical value of $2S$ and the lower critical field is absent.
We therefore need to consider
a full variational calculation in the space defined by all states $|\psi_b(i)\rangle$ and $|\psi_B(i)\rangle$ which we do next.
A preliminary discussion of results from such variational calculations formed we presented in sections~\ref{sec:pbc} and \ref{sec:hc}.

\section{Variational Approach}\label{sec:variational}
In order to develop a variational approach valid for an extended part of the phase diagram we generalize the single
defect states in Eq.~(\ref{eq:psib00}) and (\ref{eq:psib01}) to be constructed from the $|y'\rangle$ and $|x'\rangle$ states.
\begin{flalign}
&    |\psi_b(i)\rangle=|
    y'\xbond\ x'\ybond\ 
    \tcbset{colback=blue!10!white}
    \tcboxmath[size=fbox,auto outer arc, arc=5pt]{
    y'\xbond\ 
    \nearrow_i\ybond\ x'
    }
    \xbond\ y'\ybond\
    x'\xbond\ y'\ybond\
    x'\xbond\ y'\rangle,\nonumber &\\
&    |\psi_b(i)\rangle=|
    y'\xbond\ x'\ybond\ 
    y'\xbond\ 
    \tcbset{colback=blue!10!white}
    \tcboxmath[size=fbox,auto outer arc, arc=5pt]{
    x'\ybond\
    \nearrow_i\xbond\ y'
    }
    \ybond\
    x'\xbond\ y'\ybond\
    x'\xbond\ y'\rangle, \label{eq:psibp}&
\end{flalign}
transitioning from the $Y'X'$ to the $X'Y'$ pattern at bond $i$. 
As already noted, the energy cost of the ferromagnetically aligned
$ x'_i\ybond\ x' $ bond is relatively small since it occurs on a $y$-bond. Likewise for the 
$ y'_i\xbond\ y' $ bond. 
As shown in Fig.~\ref{fig:chiralZ}(a) the entanglement is very low in the soliton phase and 
we expect such product states to be of relevance.
Analogously,  we define `anti'-defects of the form
\begin{flalign}
&    |\psi_B(i)\rangle=|
    x'\xbond\ y'\ybond\ 
    \tcbset{colback=red!10!white}
    \tcboxmath[size=fbox,auto outer arc, arc=5pt]{
    x'\xbond\ 
    \nearrow_i\ybond\ y'
    }
    \xbond\ x'\ybond\
    y'\xbond\ x'\ybond\
    y'\xbond\ x'\rangle,\nonumber &\\
    &|\psi_B(i)\rangle=|
    x'\xbond\ y'\ybond\ 
    x'\xbond\ 
    \tcbset{colback=red!10!white}
    \tcboxmath[size=fbox,auto outer arc, arc=5pt]{
    y'\ybond\
    \nearrow_i\xbond\ x'
    }
    \ybond\
    y'\xbond\ x'\ybond\
    y'\xbond\ x'\rangle,\label{eq:psiBp}&
\end{flalign}
in this case transitioning from the $X'Y'$ to the $Y'X'$ pattern at bond $i$. 
As discussed, in this case the defects are now rather costly since 
since  the $y'_i\ybond\ y'$ now occurs on a $y$-bond and the $x'_i\xbond\ x'$ on a $x$-bond. The defect states, $\psi_b$ and $\psi_B$
are slight variations of the bond defects considered for the \shalf\ Kitaev chain in Ref.~\cite{Sorensen2022} and are slightly more optimal
for $S\ge 1$. However, since all such basis states are non-orthogonal the final results depend relatively little on the specific choice of basis states. 

With the states $\psi_b$ and $\psi_B$ defined we can form linear combinations of these
single defect states and perform a variational calculation within the single defect subspace. 
As illustrated in Fig.~\ref{fig:chiralZ}(a), the entanglement is very low within the soliton phase and we
therefore 
expect such linear combinations to yield very reliable results within the soliton phase.
Explicitly, we define the variational states:
\begin{equation}
    |\Psi_b\rangle=\sum_{k=1}^Na_k|\psi_b(k)\rangle,\ \ \ \ |\Psi_B\rangle=\sum_{l=2}^{N-1}{g}_l|\psi_B(l)\rangle.
    \label{eq:Psib}
\end{equation}
We refer to these states as soliton and anti-soliton states to distinguish them from the individual basis states $|\psi_b(i)\rangle$
and $|\psi_B(i)\rangle$ which we refer to as defect states or basis states. 
Correspondingly, we distinguish between soliton energies and defect energies when
referring to the energy of the linear combination and individual basis state. We also note that for $\Psi_B$ we exclude the sites $l$=1,N since 
their overlap with the lower energy $|Y'X'\rangle$ and $|X'Y'\rangle$ states is an inconvenience.

The determination of the variational coefficients, $a_k$ and $g_l$ is a straight forward optimization problem. Since the basis states
are non-orthonormal the minimum can be found by solving the generalized eigenvalue problem (see appendix~\ref{app:geneig}) in terms of the matrices
\begin{equation}
\H_{kl}=\langle \psi_b(k)|H|\psi_b(l)\rangle\ \  \mathrm{and}\ \  {\cal M}_{kl}=\langle \psi_b(k)|\psi_b(l)\rangle,\label{eq:geneig}
\end{equation}
which can be solved by standard methods. The solution of the generalized eigenvalue problem, Eq.~(\ref{eq:geneig}), determines the variational
optimized ground-states, $\Psi_b$, $\Psi_B$ in the sub-space formed by $|\psi_b(i)\rangle$ and $|\psi_B(i)\rangle$.

\begin{figure}
  \includegraphics[width=\columnwidth]{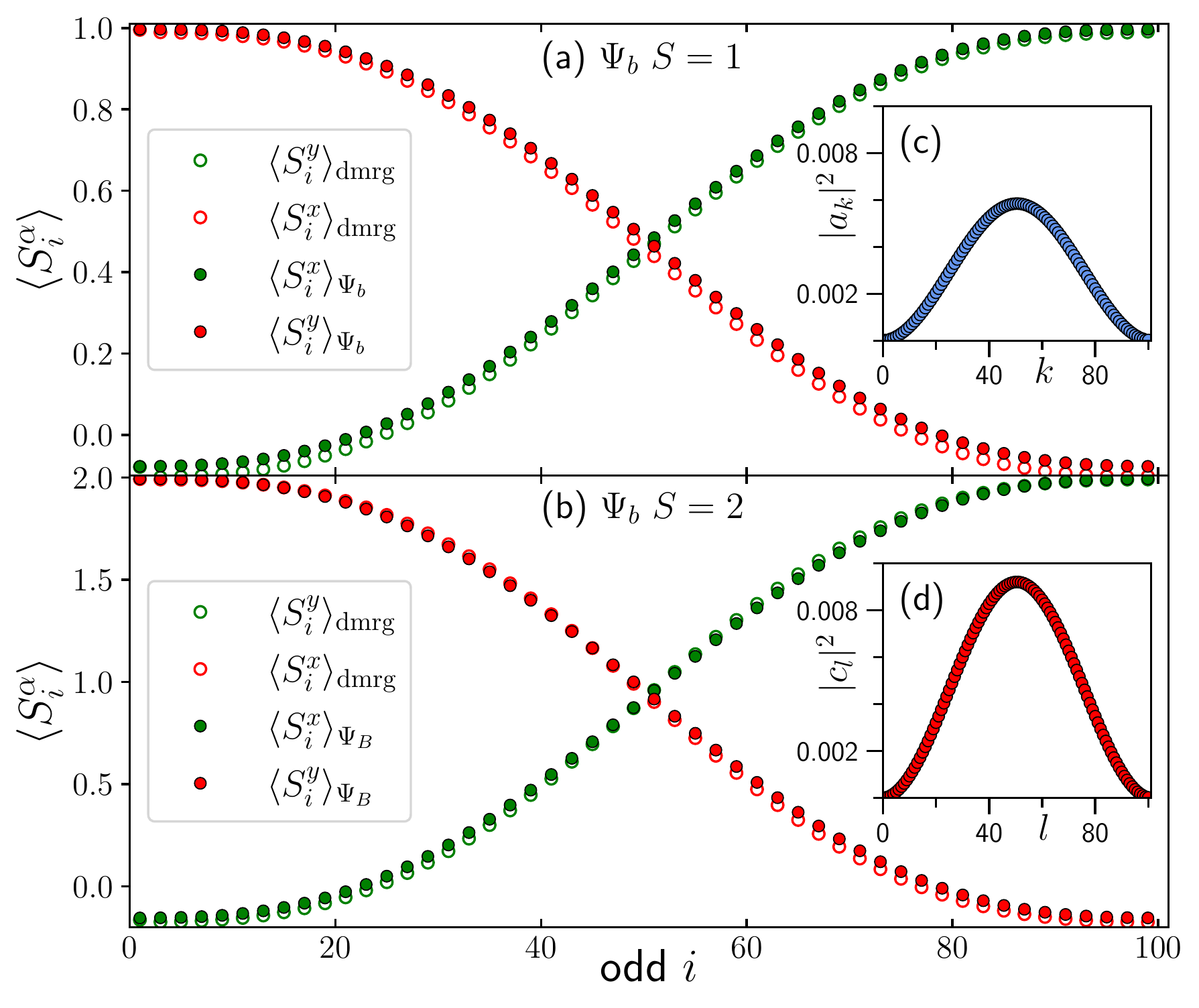}
  \caption{ $\langle S^\alpha_i\rangle$ from
   finite size DMRG results (open circles) with $N$=$100$ for the \sone\ and \stwo\ Kitaev spin chains, compared to variational results (solid circles ) for the one soliton state $\Psi_b$. To emphasize the presence of the soliton only odd sites are shown.
  (a) Results for \sone\ at $h_{xy}/K$=1.3.
  (b) Results for \stwo\ at $h_{xy}/K$=2.6.
  (c) Variational amplitudes $|a_k|^2$ for \sone.
  (d) Variational amplitudes $|c_l|^2$ for \stwo.
  }
  \label{fig:variational}
\end{figure}  

Having defined the single defect states $|\psi_b(i)\rangle$, $|\psi_B(i)\rangle$ it is straight forward to
extend the variational calculations to two-defect $bB$ states relevant for PBC by considering:
\begin{equation}
    |\psi_{bB}(i,j)\rangle=|
    \tcbset{colback=blue!10!white}
    \tcboxmath[size=fbox,auto outer arc, arc=5pt]{
    y'\xbond\ 
    \nearrow_i\ybond\ x'
    }
    \xbond\ y'\ybond\ x'\xbond 
    \tcbset{colback=red!10!white}
    \tcboxmath[size=fbox,auto outer arc, arc=5pt]{
    y'\ybond
    \nearrow_j\xbond\ x'
    }
    \ybond y'\xbond x'
    \rangle,\label{eq:psibB}
\end{equation}
and defining two-soliton states of the form:
\begin{equation}
|\Psi_{bB}\rangle=\sum_{i\neq j} a_{i,j}|\psi_{bB}(i,j)\rangle.\label{eq:2sol}
\end{equation}
Similar variational two-soliton states have previsously been considered for the $J_1$-$J_2$ \shalf\ chain~\cite{Shastry1981}
and \shalf\ Kitaev chain~\cite{Sorensen2022}. It is convenient to include the $|Y'X'\rangle$ and $|X'Y'\rangle$ states in the variational
sub-space for PBC and the variational gap to two-soliton states $\Delta_{2sol}^\mathrm{var}$ can then be directly obtained from the the eigenvalues of Eq.~(\ref{eq:geneig}). For PBC we expect $\Delta_{2sol}^\mathrm{var}$ of the spin gap, $\Delta_\mathrm{pbc}$

\subsection{Variational results for \sone\ }
\begin{figure}
  \includegraphics[width=\columnwidth]{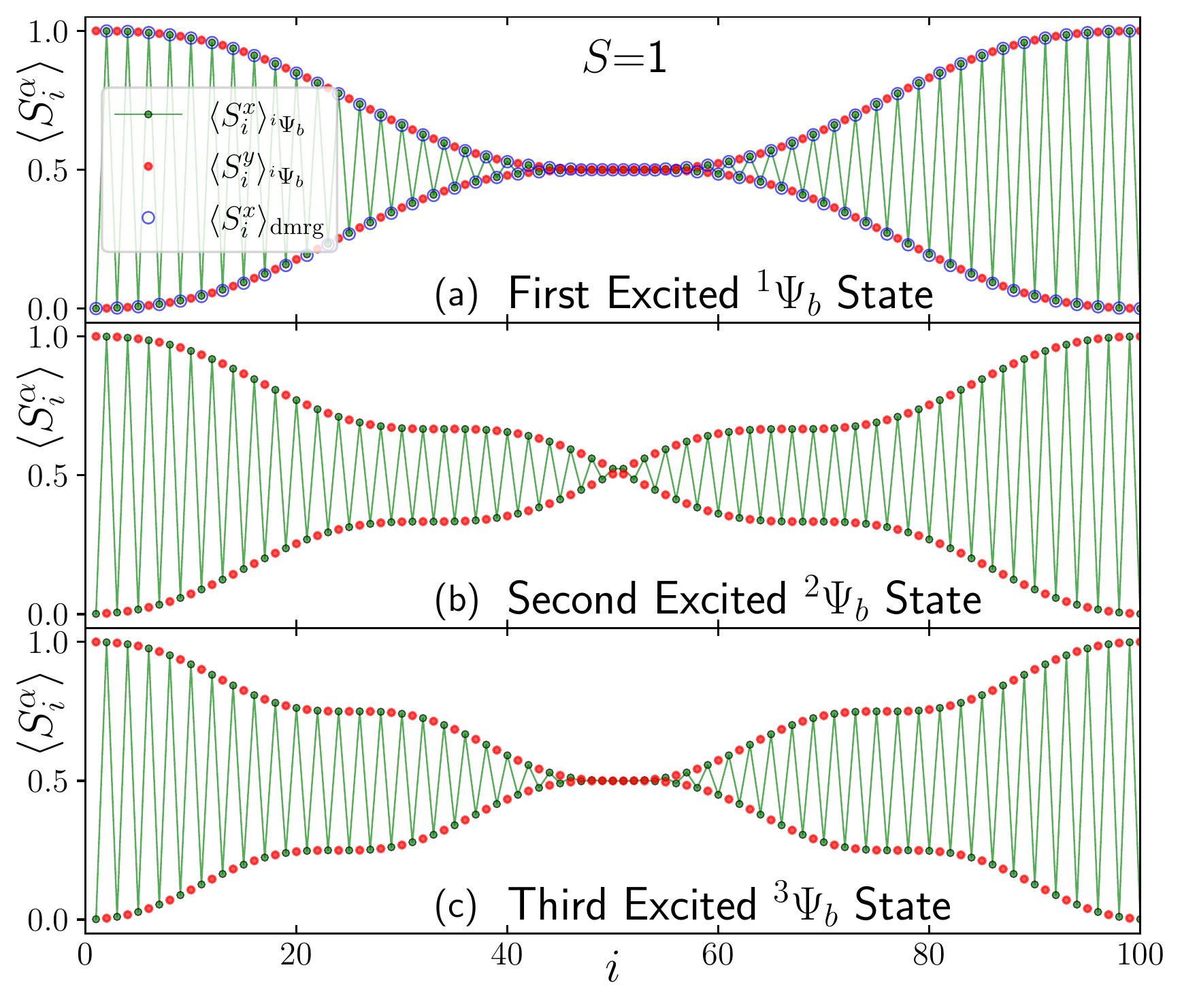}
  \caption{ $\langle S^\alpha_i\rangle$ for the \sone\ chain  at $h^\star_{xy}$ from
   variational calculations  for the {\it excited} soliton state $^n\Psi_b$.
  (a) Results for first excited state $^1\Psi_b$, compared to finite size DMRG results (open circles) for $\langle S^x_i\rangle_\mathrm{dmrg}$
  (b) Results for second excited state $^2\Psi_b$.
  (c) Results for third excited state $^3\Psi_b$.
  }
  \label{fig:ExButterflies}
\end{figure} 

\begin{table*}
\caption{\label{tab:s1}DMRG and variational, $E(\Psi_b)$, $E(\Psi_B)$, $E(Y'X')$ energies for the \sone\ chain for different field values, $h_{xy}$ and system sizes, $N$. The resulting variational estimates of $\Delta_b^\mathrm{var}$, $\Delta_B^\mathrm{var}$ and $\Delta_B^\mathrm{var}$+$\Delta_b^\mathrm{var}$ and $\Delta_{2sol}^\mathrm{var}$. These can be compared with DMRG results for $N$=1200 for $\Delta_b^\mathrm{dmrg}$ and $N$=60 for $\Delta_\mathrm{pbc}^\mathrm{dmrg}$.} 
\begin{ruledtabular} 
\begin{tabular}{c|c|c|c|c|c|c|c|c|c|c|c}
$h_{xy}$ & $N$ & $\mathrm{DMRG}$ & $E(\Psi_b)$ & {$E(\Psi_B)$} & {$E(Y'X')$} &$\Delta_b^\mathrm{dmrg}$ & {$\Delta_b^\mathrm{var}$} & {$\Delta_B^\mathrm{var}$} & {$\Delta_B^\mathrm{var}$+$\Delta_b^\mathrm{var}$} & {$\Delta^\mathrm{var}_{2sol}$} & {$\Delta_\mathrm{pbc}^\mathrm{dmrg}$}\\
\hline
\multirow{2}{*}{$h^\star_{xy}$} & 100 & -100.7453 & -100.7221 & -99. & -100. & \multirow{2}{*}{-0.7457}& -0.7221 & 1. & 0.2779 & \multirow{2}{*}{0.2788} &\multirow{2}{*}{0.2555}\\
                                & 240 & -240.7457 & -240.7225 & -239. & -240. & & -0.7225 & 1. & 0.2775 &  &\\
\hline
\multirow{2}{*}{1.3} & 100 & -93.0400 & -92.8743 & -91.2751 & -92.1725 & \multirow{2}{*}{-0.7487}& -0.7018 & 0.8974 & 0.1956 &  \multirow{2}{*}{0.2327} &\multirow{2}{*}{0.1549}\\
                     & 240 & -222.3942 & -222.0245 & -220.4254 & -221.3225 & & -0.7020 & 0.8971 & 0.1951 &  & \\
\end{tabular}
\end{ruledtabular}
\end{table*}

We first discuss our results for \sone. Representative numerical results for a few values of $h_{xy}$ and $N$ are collected in table~\ref{tab:s1}. The first check on the variational results is to directly compare the energy obtained with results from DMRG.
For \sone\ at $h^\star_{xy}$ we see that the presence of the defect lowers the energy considerably when compared to the $|Y'X'\rangle$ state
for a final result that is within $0.023\%$ ($N$=100) and $0.009\%$ ($N$=240) of the DMRG results. This is a remarkable good agreement although we note that the agreement worsens for $h_{xy}$$\neq$$h^\star_{xy}$. The agreement between $\Delta_b^\mathrm{var}$ and $\Delta_b^\mathrm{dmrg}$
is at the level of a few percent. A more detailed check on the 
variational ground-state $\Psi_b$ with OBC can be obtained by evaluating $\langle S^\alpha_i\rangle$, $\alpha$=$x,y$ and comparing to DMRG results. Variational results at $h_{xy}/K=1.3$ for the on-site magnetization (filled circles) are shown in Fig.~\ref{fig:variational}(a) where only {\it odd} sites are plotted making the change from $|y'\rangle$ on odd sites,
$|x'\rangle$ on even sites to $|y'\rangle$ on even sites, $|x'\rangle$ on odd sites, evident. The results in Fig.~\ref{fig:variational}(a) for
$h_{xy}/K=1.3$ are in excellent agreement with the DMRG results shown as open circles, with the agreement even better at $h^\star_{xy}$.
For comparison, we show results for \stwo\ in Fig.~\ref{fig:variational}(b) at $h_{xy}/K$=2.6 with equally good agreement between variational and DMRG results.

From the numerical results in table~\ref{tab:s1} it is also clear that $\Delta_B^\mathrm{var}$+$\Delta_b^\mathrm{var}$ is in good agreement
with the result, $\Delta^\mathrm{var}_{2sol}$, obtained directly from two-soliton variational calculations with Eq.~(\ref{eq:2sol}) with $N$=60. Furthermore, at $h^\star_{xy}$ both estimates are in agreement with $\Delta_\mathrm{pbc}^\mathrm{dmrg}$ obtained from DMRG calculations on periodic chains. This can be viewed as  a validation of the soliton anti-soliton picture and would indicate that interactions between 
the soliton and anti-soliton are relatively modest. However, from the discussion of the size of the soliton in section~\ref{sec:solmass} 
we expect $\xi_\mathcal{S}\sim 120$ lattice spacings in the \sone\ soliton phase, implying that much larger variational calculations will be needed to study the soliton anti-soliton interaction in detail. Regrettably, the two-soliton calculations scale as $N^2$ making such calculations numerically untractable.

\subsubsection{Excited single soliton states}
As can be seen from table~\ref{tab:s1}, in the vicinity of $h^\star_{xy}$ the spin gap for PBC is sizable, of the order $\sim 0.25 K$. It is then interesting to consider excited single soliton states~\cite{Goldstone1975,Rajaraman}. We denote such states by $^n\Psi_b$ and we can obtain reliable variational estimates for such excited states by considering the first few eigenstates when solving the generalized eigenvalue problem, Eq.~(\ref{eq:geneig}). As is clear from the results in section~\ref{sec:gaps} such excited single soliton states cause a proliferation
of low-lying levels within the soliton phase at energies below the gap for PBC. Results for $^1\Psi_b$, $^2\Psi_b$
and $^3\Psi_b$ at $h^\star_{xy}$ are shown in Fig.~\ref{fig:ExButterflies}. For the first excited state,  $^1\Psi_b$, we compare to excited state DMRG results
for $\langle S^x_i\rangle$ which are in excellent agreement with the variational results. Note that in Fig.~\ref{fig:ExButterflies} results for every site is plotted while in Fig.~\ref{fig:variational} only results for {\it odd} sites are plotted. However, in Fig.~\ref{fig:ExButterflies}
the same change from $|y'\rangle$ on odd sites,
$|x'\rangle$ on even sites to $|y'\rangle$ on even sites, $|x'\rangle$ on odd sites, occurs.

\section{Specific Heat, \sone\ }\label{sec:specheat}
\begin{figure}
  \includegraphics[width=\columnwidth]{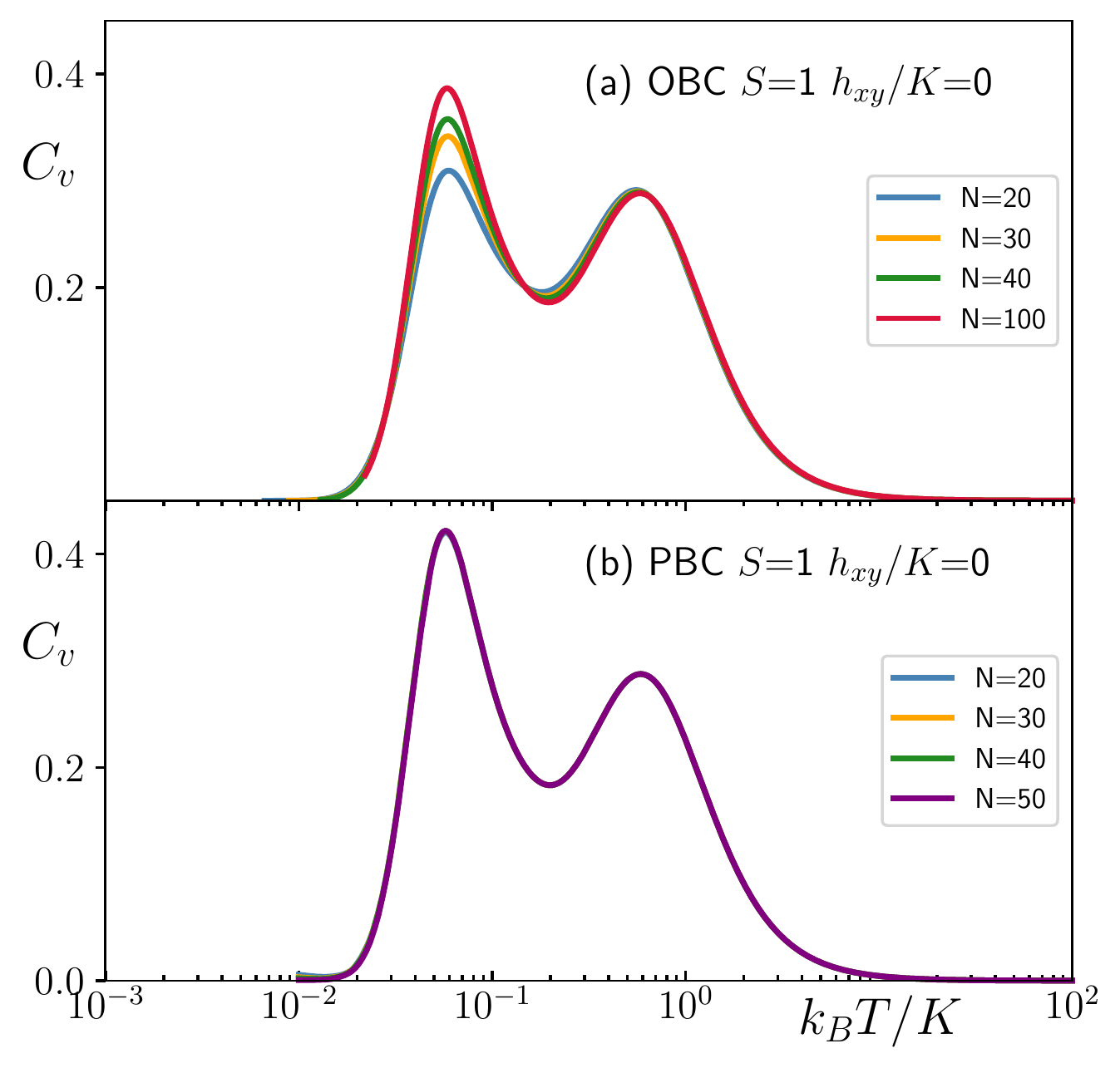}
  \caption{
  The specific heat $C_v(T)$ versus $k_BT/K$ for the \sone\ Kitaev spin chain at $h_{xy}/K$=0.0, as obtained from purification.
  (a) OBC, $N$=20,30,40,100.
  (b) PBC, $N$=20,30,40,50.
  }
  \label{fig:Cvh00}
\end{figure} 
\begin{figure}
  \includegraphics[width=\columnwidth]{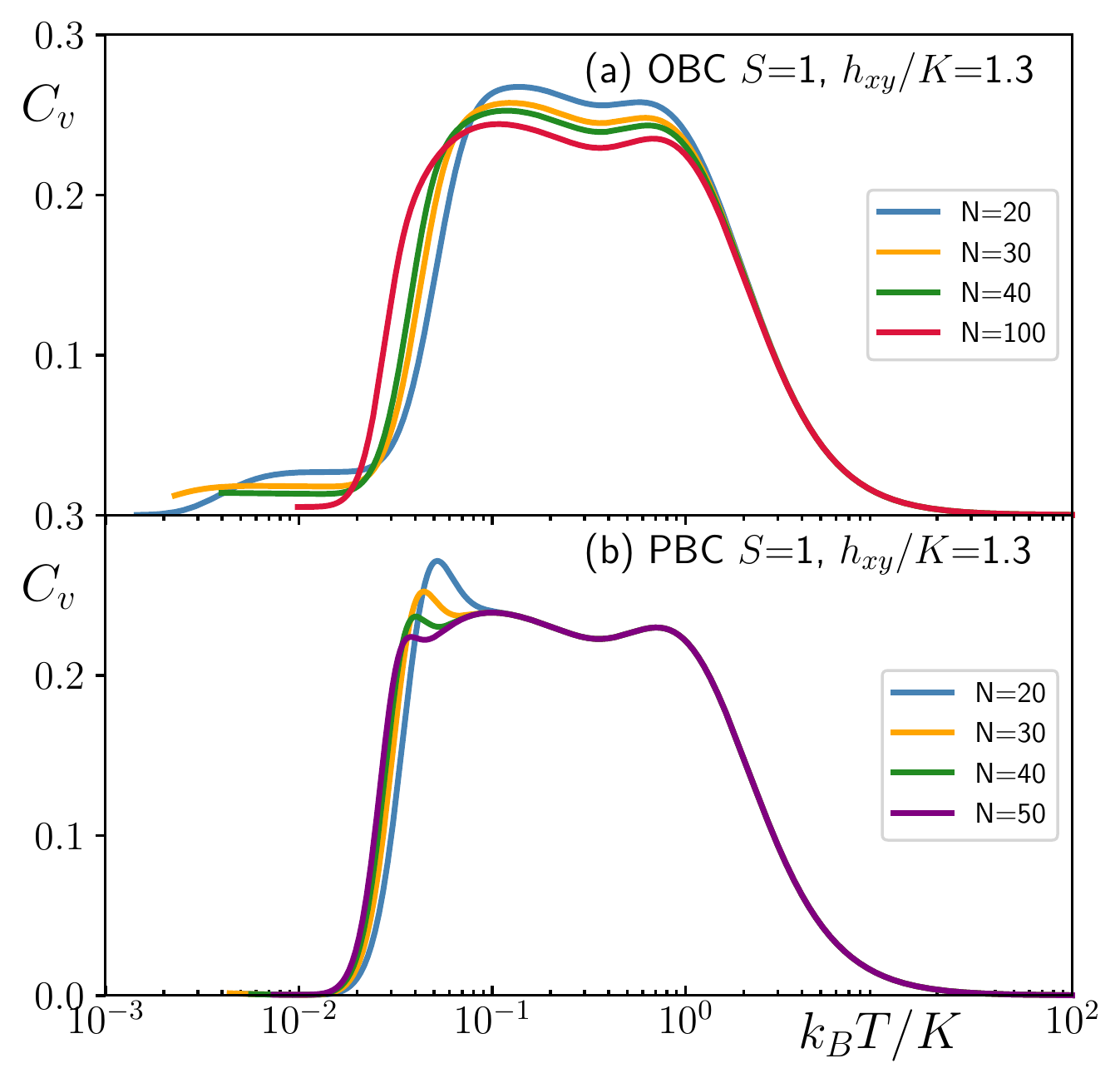}
  \caption{
  The specific heat $C_v(T)$ versus $k_BT/K$ for the \sone\ Kitaev spin chain in the middle of the soliton phase, at $h_{xy}/K$=1.3, as obtained from purification.
  (a) OBC, $N$=20,30,40,100.
  (b) PBC, $N$=20,30,40,50.
  }
  \label{fig:Cvh13}
\end{figure} 
The thermodynamics of the \sone\ Kiatev chain in zero field, $\mathbf{h}$=0, has previously been studied~\cite{Luo2021,Gordon2022} using transfer matrix 
renormalization group~\cite{Bursill1996,Wang1997} (TMRG) techniques and a perturbative effective Hamiltonian approach~\cite{Gordon2022}. To fully account for the presence of a single soliton in the low energy spectrum for OBC which breaks translational symmetry we here use a purification method
outlined in section~\ref{sec:num} that does not rely on translational symmetry. We exclusively focus on the \sone\ chain, although we expect results for other integer $S>1$ to be relatively similar.

Under periodic boundary conditions at $\mathbf{h}$=0 we show our purification results in Fig.~\ref{fig:Cvh00}(b) for the \sone\ chain for $N$=20,30,40 and 50 down to temperatures of $k_BT/K$=0.01. In complete agreement with the TMRG results from Ref.~\onlinecite{Luo2021}, finite-size effects are conspicuously absent. However, the unusual double peak structure, with peaks at $T_l/K=0.057$ and $T_h/K$=0.587 for $N$=50, of the specific heat associated with thermal fractionalization~\cite{Nasu2015,Motome2020} characteristic of Kitaev physics is clearly present arising from the separation of energy scales
as previously noted~\cite{Luo2021,Gordon2022}. The low-temperature peak has been shown to arise from excitations of the bond-parity operators, $W_l$ Eq.~(\ref{eq:bondop}),
with the average bond density, $\bar W_b$=$(1/L)\sum\langle W_l\rangle$, approaching zero at the energy scale of the low temperature peak~\cite{Luo2021,Gordon2022}.

For OBC our results at $\mathbf{h}$=0 are shown in Fig.~\ref{fig:Cvh00}(a) for $N$=20,30,40 and 100. In this case there are clearly visible finite-size effects visible in the low-temperature peak. As  the system size, $N$, is increased the low-$T$ peak increases eventually approaching the PBC result. 
We note that the results presented here for \sone\ can be straightforwardly integrated to yield the entropy. However, the results from such an integration do not show any indication of plateaus as expected to occur in the two dimensional honeycomb models~\cite{Oitmaa2018}.

The results in Fig.~\ref{fig:Cvh00} should be contrasted with the results in Fig.~\ref{fig:Cvh13} obtained for the \sone\ chain close to the center of the soliton phase at $h_{xy}/K$=1.3. Compared to the $\mathbf{h}$=0 results the first observation is that the separation of energy scales present at $\mathbf{h}$=0 inducing the double peak structure is now  significantly reduced and replaced with an almost constant specific heat between temperatures of $k_BT/K\sim 0.05$ although several not very well defined peaks are visible. 
For PBC, Fig.~\ref{fig:Cvh13}(b) it is possible to locate 3 peaks, two of which are almost independent of $N$, however, the lowest temperature
peak dramatically decreases with increasing system size with significant weight in $C_v$ shifting to lower temperatures. 
It is natural to associate this lowest temperature peak with the $bB$ soliton states.
Within the picture
we have been proposing here, where the spin gap for PBC, $\Delta_\mathrm{pbc}$ in the soliton phase arises from the presence of 
such $bB$ states with
both a soliton and an anti-soliton, it is natural to expect rather pronounced finite-size effects due to the significant size of the solitons, $\xi_\mathcal{S}\sim 120$ lattice spacings in the \sone\ soliton phase. 
This would explain the strong size dependence of the peak. We expect a continuum of such $bB$ states starting above the spin gap which is consistent with the results for PBC in Fig.~\ref{fig:Cvh13}(b). From the results in Fig.~\ref{fig:gaps}(b), we note that $\Delta_\mathrm{pbc}\sim0.1548/K$ at $h_{xy}/K$=1.3 whereas the low-$T$ peak for $N$=50 occurs at $k_BT/K=0.038$ implying a significant density of states starting at $\Delta_\mathrm{pbc}$.

The more interesting features of the specific heat are observed for OBC, where we show results in Fig.~\ref{fig:Cvh13}(a) at $h_{xy}/K$=1.3 for
$N$=20,30,40 and 100.
For OBC the finite-size effects are now pronounced for any $k_BT/K<1$. It is natural to view this observation as being due to a considerable spatial size of the excitations responsible dor the energy fluctuations. Most strikingly, for temperatures below $k_BT/K\sim 0.02-0.03$ a 'foot' of the specific heat can be observed with $C_v$ almost constant over a considerable range of temperatures, albeit at a very low value. The value
of $C_v$ over this plateau appears to be decreasing with $N$. Unfortunately, due to size and temperature limitations it has not been possible to perform calculations at larger $N$, lower $T$. Since this 'foot' in $C_v$ is only present for OBC at temperatures lower than for PBC it is clear that it most arise from excitations only present with OBC. We therefore ascribe this feature to the single soliton ground-state for OBC, excitations of which (Fig.~\ref{fig:ExButterflies}) should significantly contribute to $C_v$ at energies below $\Delta_\mathrm{pbc}$

\section{Discussion}\label{sec:Conclusion}
Here we discuss a few open questions and future directions.
The variational picture of the soliton phase that we have been advocating here rely on the presence of a gap for periodic boundary conditions within the soliton phase. At the special point $h^\star_{xy}$, the $|YX\rangle$ and $|XY\rangle$ product states are exact ground-states.
It therefore seems plausible that an analytic proof of a gap at $h^\star_{xy}$ can be established.
So far we have not been able to develop such a proof due to the low
symmetry at $h^\star_{xy}$ and the degeneracy of the ground-state with PBC in the soliton phase.

Under open boundary conditions we have shown here that the ground-state for any $N$ always contain a single soliton which can exist in excited states leading to the formation of in-gap states. Excited states of quantum solitons have been considered before~\cite{Rajaraman,Vachaspati} and are usually associated with a discrete harmonic oscillator like spectrum. In the present
case it is not clear if the in-gap states created by excitations of the soliton form a continuous band or if they form discrete states in the thermodynamic limit. The energy of the lowest excited states appear to approach the ground-state quickly as $N$ is increased
but due to limitations in the size of the systems we can reliably study it has not been possible to determine if they indeed become degenerate with the ground-state in the thermodynamic limit. The degeneracy of the ground-state with OBC is hence an open question. We leave both these questions for further study.

As illustrated in Fig.~\ref{fig:phasediagram} the size of the soliton islands grow with increasing $S$ and one might ask the question what happens in the  $S\to\infty$ classical limit. Classical Monte Carlo simulations are inconclusive in the low field limit but one might speculate that the soliton phase would occopy the entire phase diagram for any $|h|<2S$ but so far we have not been able to establish a proof of this.

It would be of considerable interest to identify realistic low-dimensional Kitaev materials
to test the soliton physics presented here. 
Recently it was proposed that CoNb$_2$O$_6$ exhibits signatures of Kitaev physics known as twisted Kitaev chain~\cite{Morris2021}, albeit with \shalf\ FM Kitaev interaction and hence not the AFM Kitaev interaction required for our scenario.
However, it seems likely that 
the AFM Kitaev interaction required for the soliton phase can occur in \sone\ systems. 
Note that the effective \shalf\  Kitaev materials with $d^5$ have a predominantly FM Kitaev interaction as the inter-orbital exchange process among $t_{2g}$-orbitals leads to a FM Kitaev interaction\cite{jk2009prl,rau2014prl}. On the other hand, in \sone\ systems with $d^8$, the Kitaev interaction is AFM  as found from the exchange processes of $e_g$-orbitals via strong spin-orbit coupling at anions~\cite{Stavropoulos2019}.
It was also suggested that $4f^1$ system contains AFM Kitaev interaction due to the spatial anisotropy of the f orbitals and the small crystal field splitting.\cite{Motome2020Materials}
Thus, the soliton phase occuring in the AFM Kitaev interaction under the magnetic field can be investigated if solid-state materials with
quasi-one-dimensional $d^8$ systems and edge sharing heavy ligands or $4f^1$ can be identified.
Since the solitons we have discussed here  are particularly well defined for large integer spin, if such low-dimensional
AFM Kitaev materials with large $S$ can be found, it would offer the best possibility for observing the solitons.
Finally, we remark that it would interesting to study the dynamics of the solitons in a non-equilibrium setting.

\begin{acknowledgments}
This research was supported by NSERC and CIFAR 
and was enabled in part by support provided by SHARCNET (sharcnet.ca) and the Digital Research Alliance of Canada (alliancecan.ca).
Part of the numerical
calculations were performed using the ITensor library~\cite{itensor}.
\end{acknowledgments}

\appendix
\section{The Generalized Eigenvalue Problem}\label{app:geneig}
Let us consider a set of states $\{|b_i\rangle\}_{i=1}^N$ and a Hamiltonian $\H$.  We can expand a generic state $|\psi\rangle$ on such basis
states writing
\begin{equation}
|\psi\rangle=\sum_{i=1}^N c_i|b_i\rangle
\end{equation}
According to the variational
principle the minimum condition is then written as the generalized eigenvalue problem
\begin{equation}
\sum_j\left(\H_{ij}-E \mathcal{M}_{ij}\right)c_j=0,  \label{eq:a2}
\end{equation}
where $\H_{ij}$=$\langle b_i|\H|b_j\rangle$ and $\mathcal{M}_{ij}$ is the overlap matrix $\langle b_i|b_j\rangle$. In the case where
$\langle b_i|b_j\rangle$=$\delta_{ij}$ this reduces to the standard eigenvalue problem. We can write Eq.~(\ref{eq:a2}) in matrix form
as
\begin{equation}
    \H \mathbf{c}=E\mathcal{M}\mathbf{c},\label{eq:a3}
\end{equation}
which defines a generalized eigenvalue problem.
To solve Eq.~(\ref{eq:a3}) we first solve the standard eigenvalue problem
\begin{equation}
    \mathcal{M}\mathbf{d}=m\mathbf{d}.
\end{equation}
If the states $\{|b_i\rangle\}_{i=1}^N$ are linearly independent then $\mathcal{M}$ is positive definite and hermitian which implies we
can find a unitary matrix $D$ such that $D^\dagger\mathcal{M}D$ is a diagonal matrix. Since all $m>0$ we can then define
\begin{equation}
A_{ij}\equiv \frac{D_{ij}}{\sqrt{m_j}}
\end{equation}
so that $A^\dagger\mathcal{M}A$=$I$. If we now define
\begin{equation}
    \mathbf{c}=A\mathbf{v},
\end{equation}
then Eq.~(\ref{eq:a3}) can be written as
\begin{equation}
    \H A\mathbf{v}=E\mathcal{M}A\mathbf{v}.\label{eq:a6}
\end{equation}
If we now apply the matrix $A^\dagger$ from the left we then obtain
\begin{equation}
   A^\dagger \H A\mathbf{v}=EA^\dagger \mathcal{M}A\mathbf{v}=E\mathbf{v},\label{eq:a7}
\end{equation}
which is now a standard eigenvalue problem for the matrix $A^\dagger \H A$. We have then reduced the solution of
the generalized eigenvalue problem to the solution of two standard eigenvalue problems.
\bibliography{references}
\end{document}